\documentclass[preprint2,twocolappendix]{aastex6}
\usepackage[T1]{fontenc}
\usepackage{microtype}
\usepackage{afterpage,amsmath,amssymb,graphicx,natbib, color}
\usepackage{hyperref}
\usepackage{dcolumn}
\newcolumntype{d}[1]{D{.}{.}{#1}}

\usepackage{array}   % for \newcolumntype macro
\newcolumntype{L}{>{$}l<{$}} % math-mode version of ``l'' column type
\newcolumntype{R}{>{$}r<{$}}
\newcolumntype{C}{>{$}c<{$}}
\newcommand\plotoneman[2]{\centering \leavevmode
\includegraphics[width=#2\linewidth]{#1}}

\newcommand\citeeg[1]{\citep[e.g.,][]{#1}}
\newcommand\lsim{\lesssim}
\newcommand\gsim{\gtrsim}

\newcommand\power[2]{\ensuremath{#1\times10^{#2}}}
\newcommand\Msun{\ensuremath{\mathrm{M}_\odot}}

\newcommand\Reff{\ensuremath{\mathrm{R}_e}}
\newcommand\Mstar{\ensuremath{\mathrm{M}_\star}}
\newcommand\mgrad{\ensuremath{\nabla\mathrm{[Z/H]}}}
\newcommand\agrad{\ensuremath{\nabla\mathrm{age}}}
\newcommand\lumgrad{\ensuremath{\nabla\Sigma_V}}
\newcommand\cgrad{\ensuremath{\nabla\textrm{g-r}}}
\newcommand\fex{\ensuremath{\mathrm{f}_\mathrm{ex}}}
\newcommand\FGM{\ensuremath{\mathrm{F}(\mathrm{G},\mathrm{M}_{20})}}

\newcommand\changed[1]{{\bf #1}}
\renewcommand\changed[1]{#1}

\begin{document}

\title{The information content of stellar halos: Stellar population
  gradients and accretion histories in early-type Illustris galaxies}

\author{B.~A.~Cook\altaffilmark{1}, C.~Conroy\altaffilmark{1}, A.~Pillepich\altaffilmark{1,2},
  V.~Rodriguez-Gomez\altaffilmark{1,3}, and L.~Hernquist\altaffilmark{1}}
\altaffiltext{1}{Harvard-Smithsonian Center for Astrophysics, 60 Garden St., Cambridge, MA 02138, USA}
\altaffiltext{2}{Max-Planck-Institut f\"ur Astronomie, K\"onigstuhl 17, 69117 Heidelberg, Germany}
\altaffiltext{3}{Department of Physics \& Astronomy, Johns Hopkins
  University, 3400 N.~Charles Street, Baltimore, MD 21218, USA}
\email{bcook@cfa.harvard.edu}

\slugcomment{Submitted: \today} \shorttitle{Information Content in
  Illustris Stellar Halos} \shortauthors{Cook \textit{et al.}}

\begin{abstract}
Long dynamical timescales in the outskirts of galaxies preserve the
information content of their accretion histories, for example in the
form of stellar population gradients. We present a detailed analysis
of the stellar halo properties of a statistically representative
sample of early-type galaxies from the Illustris simulation and show
that stellar population gradients at large radii can indeed be used to
infer basic properties of galactic accretion histories. We measure
metallicity, age, and surface-brightness profiles in quiescent
Illustris galaxies ranging from $\Mstar = 10^{10} -
2\times10^{12}\Msun$ and show that they are in reasonable agreement
with observations. At fixed mass, galaxies that accreted little of
their stellar halo material tend to have steeper metallicity and
surface-brightness profiles between $2-4$ effective radii ($\Reff$)
than those with larger accreted fractions. Profiles of metallicity and
surface-brightness in the stellar halo typically flatten from $z=1$ to
the present. This suggests that the accretion of stars into the
stellar halo tends to flatten metallicity and surface-brightness
profiles, a picture which is supported by the tight correlation
between the two gradients in the stellar halo. \changed{We find no
  statistical evidence of additional information content related to
  accretion histories in stellar halo metallicity
  profiles beyond what is contained in surface-brightness
  profiles}. Age gradients in the stellar halo do not appear to be
sensitive to galactic accretion histories, and none of the stellar
population gradients studied are strongly correlated with the mean
merger mass-ratio. Future observations that reach large radii outside
galaxies will have the best potential to constrain galactic accretion
histories.
\end{abstract}


%======================================================
\section{Introduction}
\label{s.Intro}
The $\Lambda$CDM model of cosmology makes strong predictions regarding
the expansion of the universe and the hierarchical growth of structure
on large scales that have been supported by numerous observations
\citep{Smoot1992,Eisenstein2005,Hinshaw2013}, but the formation and
evolution of galaxies remain poorly understood \citep[see reviews
  in][]{Conselice2014,Somerville2015}. In particular, the assembly
history of early-type galaxies (ETGs) represents a major unsolved
problem: how did the red-and-dead, elliptical galaxies we see today
form, and what processes drove their evolution from early times
\citep[see reviews in][]{Renzini2006,Kormendy2009,Cappellari2016}?

Early-type galaxies host predominantly old populations of stars that
are thought to have formed on short timescales ($\approx 1
\mathrm{Gyr}$)
\citeeg{Worthey1994,Trager2000,Thomas2005,Conroy2014b}. In some
respects these observations offer support for the classic
monolithic-like collapse scenario
\citeeg{Eggen1962,Partridge1967,Larson1975,Carlberg1984} that
quiescent galaxies formed at high redshifts from the gravitational
collapse of massive primordial gas clouds resulting in brief, intense
bursts of star formation that were quickly quenched. However, it is
now generally accepted that galaxies form through a combination of two
complementary mechanisms: the \textit{in-situ} formation of new stars,
and the accretion of \textit{ex-situ} stars via mergers
\citeeg{Kobayashi2004,Oser2010,Pillepich2015,Rodriguez-Gomez2016}.

These two formation channels can, for example, help explain the
dramatic size evolution of massive early-type galaxies: extremely
compact ``red nuggets'' are found at $z\approx2$ that are factors of
$2-4$ times smaller than $z=0$ ETGs of similar masses
\citeeg{Daddi2005,Trujillo2006,Trujillo2007,VanDokkum2008}. Since
then, these galaxies grew largely through accretion of material in
their outskirts \citep{VanDokkum2010,VanderWel2014,Wellons2016}. The
accumulated light at large radii around ETGs should thus contain
important information about their accretion histories.

In-situ and ex-situ formation are also predicted to leave different
signatures in the radial gradients of stellar populations
(metallicity, age, abundance ratios) within early-type galaxies. When
stars form in-situ, the lower gas densities in the outskirts of
galaxies will result in lower star formation rates, producing fewer,
less metal-enriched stars than the galaxy interiors. This effect is
enhanced by stellar winds and other feedback mechanisms, which can
more efficiently remove metals from the outer galactic regions than
from the deep potential wells at the centers
\citep{Carlberg1984,Kobayashi2004}. In-situ formation of stars is
therefore predicted to imprint steeply negative metallicity and
surface-brightness gradients. In contrast, accretion
and mergers will deposit significant amounts of tidally-stripped stars
into galactic outskirts. Low-mass satellites will tend to bring in
more metal-poor stars and be tidally stripped at larger radii than
will massive galaxies, such that the resulting effects of this
ex-situ growth on metallicity and surface-brightness profiles could
depend on the typical merger ratio. In practice, it is believed that
the net effect of realistic merger histories tends to flatten both
profiles \citeeg{Bekki1999,DiMatteo2009,Font2011}.

$\Lambda$CDM predicts a wide variety of accretion histories due to the
stochastic nature of mergers but also predicts systematic trends with
mass \citep{Lacey1994a}. Observations of gradients in stellar
properties (including metallicity and surface brightness) offer
prospects to distinguish between in-situ and ex-situ mass growth. The
information content of mergers should be largely preserved at the
present epoch at large radii around galaxies due to long dynamical
timescales \citeeg{Eggen1962,Bullock2005}, making galaxy outskirts the
ideal regions for probing accretion histories.

Significant progress has been made in observing structures known as
\textit{stellar halos}: diffuse light seen in the outskirts around
both early- and late-type galaxies that is believed to be the remnant
of this hierarchical accretion. The Milky Way stellar halo has been
thoroughly studied through resolved star counts
\citeeg{Ibata1994,Helmi1999,Ivezic2000,Majewski2003,Bell2008a} as has
that of our neighbor M31
\citeeg{Ibata2001,McConnachie2009,Gilbert2014,Gregersen2015}.  Both
stellar halos contain significant substructure, including streams from
satellites that were recently devoured.

Characterizing the stellar halos around more distant galaxies is
challenging, but the field is progressing rapidly. Photometry is used
to measure the integrated light from stellar halos, either on an
individual basis
\citeeg{Mihos2005,Martinez-Delgado2010,Martinez-Delgado2012,VanDokkum2014,Duc2014,Buitrago2016,Huang2016a}
or in stacked images \citeeg{Tal2011,LaBarbera2012,DSouza2014}.
Spectroscopic metallicity gradients are beginning to reach large radii
\citeeg{Sanchez-Blazquez2007,Foster2009,Spolaor2010b,Coccato2010,Coccato2011},
a measurement that has been greatly aided by the introduction of
multi-object spectrographs \citeeg{Pastorello2014,Pastorello2015} and
integral-field units
\citeeg{Kuntschner2010,Greene2012,Greene2015,GonzalezDelgado2015,Oliva-Altamirano2015},
although this remains an expensive measurement, particularly given the
low surface-brightnesses and large field-of-view required.

The increasing richness of stellar halo observations necessitates
detailed model predictions in order to link the properties of an
individual stellar halo to its accretion history. Unfortunately, very
high resolution is required to reliably model the diffuse stellar halo
regions. One approach to overcoming this barrier is to focus on the
gravitational dynamics of particles, either with analytical models for
the host galaxy potential
\citeeg{Johnston1996,Bullock2005,Amorisco2015,Amorisco2016a} or using
a ``particle tagging'' technique applied to N-body cosmological
simulations \citep{Cooper2010,Helmi2011,Lowing2014a}. Both approaches
make significant simplifying assumptions \citep[see][]{Bailin2014} but
are able to reach very high resolution and explore important dynamical
effects in a controlled environment. Hydrodynamical simulations
focusing on stellar halos are able to self-consistently grow realistic
galaxies and stellar populations, but have previously been restricted
to relatively small sample sizes
\citeeg{Abadi2006,Tortora2011,Font2011,Hirschmann2015,Cooper2015,Pillepich2015}
that may not reproduce the large diversity of galactic accretion
histories.

In this paper, we present analysis of the stellar population gradients
in early-type galaxies from the Illustris hydrodynamical simulation
\citep{Vogelsberger2014a}. Illustris is the first cosmological
simulation that is simultaneously able to a) probe the wide variety
of galactic accretion histories in a $\Lambda$CDM cosmology because of
its large volume and sample size, b) generate realistic galaxies in
this cosmology using a self-consistent galaxy formation model, and c)
produce a large statistical sample of well-resolved stellar halos
around these galaxies. Previous work by \citet{Pillepich2014}
quantified the relation between the stellar mass profile in Illustris
stellar halos and the formation history of their underlying dark
matter halos, including halo formation time, time from the last major
merger, and fraction of stars accreted from infalling satellites and
mergers. We expand this analysis to consider the prospects for
connecting observable stellar population gradients to the accretion
histories of stellar halos and the redshift-evolution of the stellar
population profiles. In particular, we investigate whether there is
additional information content in stellar halo metallicity and age
profiles relative to that retained in the profile of stellar mass (or
surface-brightness), as studied in \citet{Pillepich2014}.

This work is organized as follows. We begin by describing the details
of the Illustris simulation (\S\ref{s.Methods.Illustris}) and our
sample of quiescent galaxies (\S\ref{s.Methods.sample}). We then
define our methods used to measure stellar population gradients
(\S\ref{s.Methods.gradients}) and to quantify accretion histories
(\S\ref{s.Methods.histories}). In \S\ref{s.Results}, we compare the
stellar population gradients measured in Illustris ETGs to available
observations, present our findings that gradients in the stellar halo
are sensitive to the halo's assembly history, and discuss the
redshift-evolution of stellar population profiles. In
\S\ref{s.Discussion}, we discuss our results regarding of the
information content retained within stellar halos, and we summarize in
\S\ref{s.Summary}.

%======================================================
\section{Methods}
\label{s.Methods}
%-----------------------
\subsection{The Illustris simulation}
\label{s.Methods.Illustris}

All analysis in this work is based on galaxies from the Illustris
simulation. Detailed descriptions can be found in
\citet{Vogelsberger2014a}, \citet{Vogelsberger2014b}, and
\citet{Genel2014a}. The simulation data has been released to the
public \citep{Nelson2015a}, and is available online
(\href{http://www.illustris-project.org}{http://www.illustris-project.org}). We
summarize the relevant properties of the simulation below.

Illustris is a suite of hydrodynamical cosmological simulations
(periodic box, 106.5 Mpc on a side) using the adaptive mesh code {\sc
  Arepo} \citep{Springel2010} and including runs at multiple
resolutions. Its galaxy formation model implements the most important
physical processes, including stellar formation and feedback, chemical
enrichment, radiative cooling, and feedback from AGN. The most
highly-resolved run (Illustris-1, hereafter simply Illustris) has a
mass resolution of $m_{DM} = \power{6.26}{6} \Msun$ for dark matter
and $m_{baryon} \approx \power{1.26}{6} \Msun$ for the baryonic
component. At $z=0$, gravitational forces for dark matter and baryons
are resolved to softening lengths of 1.4 kpc and 0.7 kpc,
respectively. The simulations were run from $z=127$ to $z=0$ using
cosmological parameters consistent with \textit{WMAP9}
\citep[$h=0.704, \Omega_\Lambda=0.7274,
  \Omega_m=0.2726$;][]{Hinshaw2013}.

Halos and galaxies in Illustris are identified using the
Friends-of-Friends (FOF) and {\sc Subfind} algorithms
\citep{Springel2001,Dolag2009}. Throughout this work, we study the
stellar components of central galaxies, which are defined as the most
massive {\sc Subfind} subhalos within a given FOF halo (i.e.~not
satellites/subhalos of a more massive galaxy). Stellar particles are
removed from the central galaxies if they are gravitationally bound to
a satellite subhalo.

At $z=0$, the Illustris volume contains over 40,000 galaxies resolved
with at least 500 stellar particles. Illustris has been shown to
reproduce basic observational properties including the galaxy stellar
mass function and the evolution of cosmic star formation
\citep{Genel2014a} as well as a reasonable diversity of morphologies
and colors, including early- and late-type galaxies
\citep{Torrey2015a,Snyder2015b}. The most massive central galaxies
reproduced in the simulation have stellar masses (\Mstar, defined
hereafter as the stellar mass within twice the stellar half-mass
radius) around $\power{2}{12}\Msun$. In \citet{Pillepich2014},
Illustris galaxies were shown to have well-resolved stellar halos.

There are known issues with Illustris' galaxy formation model. Due to
limitations in the implementation of stellar and AGN feedback, the
cosmic star formation rate density is too high at $z\lsim1$, which may
be responsible for quiescent galaxies (the red sequence) being
somewhat underrepresented at $z=0$ \citep{Vogelsberger2014a}. The
physical sizes of galaxies with stellar masses $\Mstar \lsim
10^{10.7}\Msun$ are found to be too large by factors of a few
\citep{Snyder2015b}. Additionally, due to a numerical implementation
choice, a very small fraction of stellar particles have
unrealistically low metallicities ($\mathrm{[Z/H]} < -10$), and were
thus removed from our analysis. We direct the reader to
\citet{Nelson2015a} for more detail on each of these issues, and we
discuss the possible impacts on our conclusions in
\S\ref{s.Discussion}.

%-----------------------
\subsection{A sample of quiescent galaxies and their properties}
\label{s.Methods.sample}

\begin{figure*}
  \plotoneman{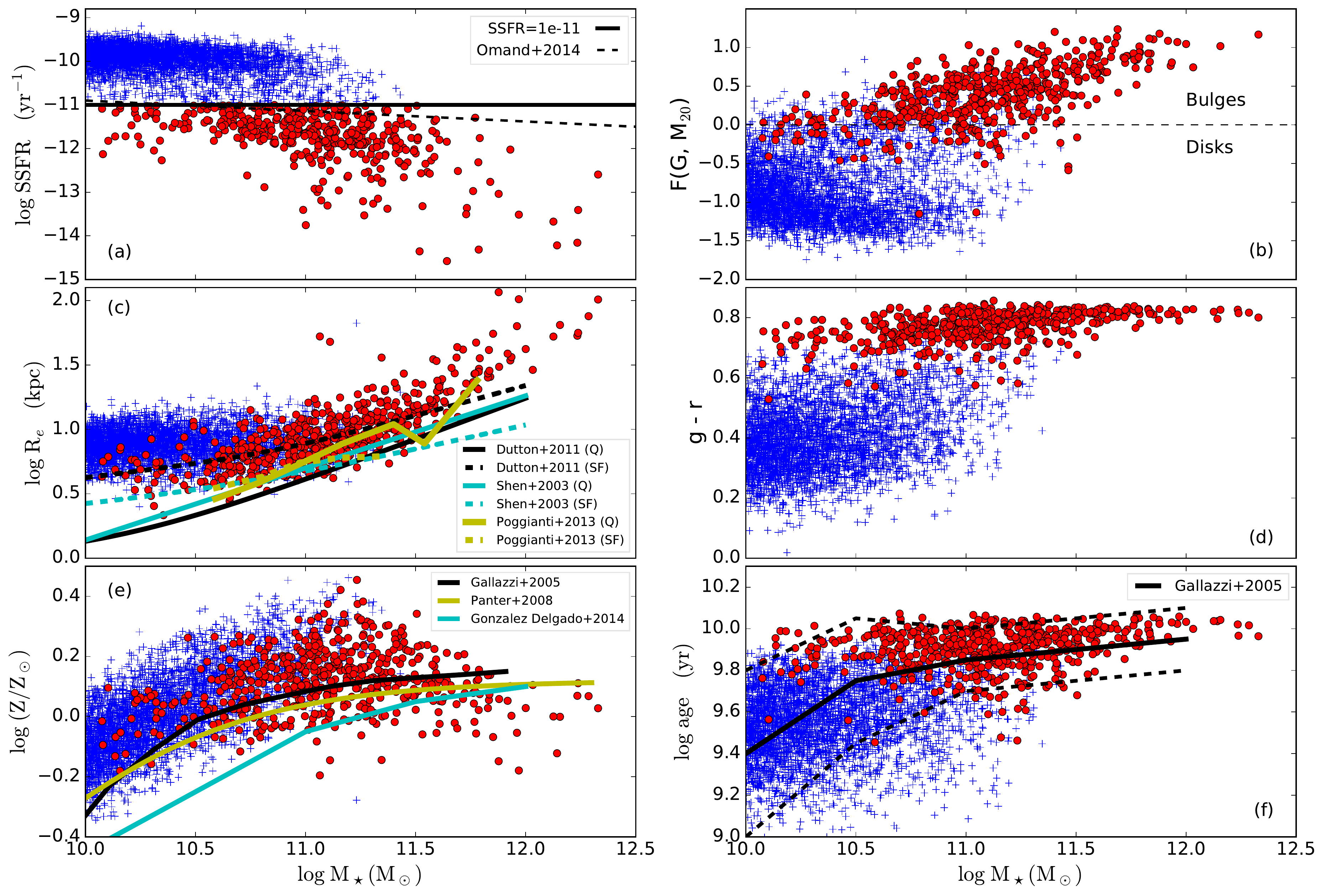}{0.995}
  \caption{Comparisons of our quiescent Illustris galaxy sample to
    observations regarding several common relations. References marked
    with Q (SF) refer to samples of early-type (late-type)
    galaxies. The sample is comprised of 537 ``quiescent'' Illustris
    galaxies. Red circles (blue crosses) show quiescent (star-forming)
    central galaxies. \changed{(a)} The distribution of specific
    star-formation rates. Quiescent galaxies are defined as SSFR $\leq
    10^{-11}$yr$^{-1}$, which selects galaxies below the star-forming
    main-sequence and agrees with the observed bimodality in SDSS
    galaxies \citep{Omand2014}. \changed{(b) The distribution of morphologies,
    parametrized by \FGM, a combination of the Gini and
    M$_{20}$ coefficients \citep[see][]{Snyder2015b}. Nearly all of the
    quiescent galaxies are bulge-dominated ($\FGM>0$). (c)} The sizes of
    Illustris galaxies (\Reff) below $\Mstar \lsim 10^{10.7} \Msun$
    are known to be too large \citep{Snyder2015b} when compared to the
    observed size-mass relation
    \citeeg{Shen2003,Dutton2011,Poggianti2013}. The large radii
    ($\gsim 30$kpc) found at high masses are likely due to the ICL
    component, which we do not remove. \changed{(d) The distribution of $g-r$
    colors, measured within $\frac{1}{2}\Reff$, shows that the
    quiescent galaxies all have red colors, as is to be expected from
    their low star-formation rates. (e)} The
    mass-metallicity relation for Illustris galaxies, in comparison to
    observations
    \citep{Gallazzi2005,Panter2008,GonzalezDelgado2014,Conroy2014b}. Metallicities
    are luminosity-weighted (V-band) and measured within
    $\frac{1}{2}\Reff$. The low metallicities at high masses are
    likely an artifact of pollution by the ICL (see
    text). \changed{(f)} The mass-age relation (V-band
    weighted, measured within $\frac{1}{2}\Reff$) among Illustris
    galaxies is compared to observations
    \citep{Gallazzi2005,Conroy2014b}. The quiescent sample is made up
    of old stellar populations.}
  \label{fig.sample}
\end{figure*}

Our sample (summarized in Figure \ref{fig.sample}) is comprised of
Illustris central galaxies with stellar mass $\Mstar \geq
10^{10}\Msun$ at $z=0$. This selection criterion ensures that there
are enough stellar particles to resolve the outskirts of the galaxies
(at least 1000 particles beyond $2\;\Reff$). Of these, we study
\textit{quiescent} galaxies, which we define as having an
instantaneous specific star-formation rate (in gas cells within twice
the stellar half-mass radius) of $\mathrm{SSFR}\leq
10^{-11}\mathrm{yr}^{-1}$. This selects red-and-dead galaxies that are
below the star-forming main sequence (SSFR $\approx
10^{-10}$yr$^{-1}$) and can be compared to observational samples of
ETGs.

The final sample includes 537 quiescent galaxies, with stellar masses
ranging from $10^{10}$ to $\power{2}{12}\Msun$. Most (491) of the
galaxies are between $10^{10.5} \leq \Mstar \leq 10^{12}$;
more-massive galaxies are rare in the Illustris volume, and
less-massive galaxies are mostly star-forming. In \changed{panel (a)}
of Figure \ref{fig.sample}, we show that our ``quiescent'' selection
criterion of $\mathrm{SSFR}\leq 10^{-11}\mathrm{yr}^{-1}$ is in good
agreement with the bimodality observed in SDSS galaxies, as described
in \citet{Omand2014}.

We measure galaxy sizes using the V-band light profiles, with luminosities
calculated via the stellar population synthesis model of
\citet{Bruzual2003}. The positions of each particle are projected
against a random line-of-sight (see Appendix \ref{a.projection} for
more details on the important effects of projection) to determine each
galaxy's effective radius \Reff, defined as the radius that contains
one half the total light.

\changed{In panel (b) of Figure \ref{fig.sample}, we show the
  distribution of morphologies in the Illustris galaxies, as
  calculated in \citet{Snyder2015b}. Morphologies are measured in
  terms of Gini's coefficient (G) and M$_{20}$, two non-parametric
  measurements of a galaxy's spatial distribution. Disk- and
  bulge-dominated galaxies are found to lie in different regions of
  $\mathrm{G}-\mathrm{M}_{20}$ space \citep{Lotz2004}. The two
  measurements are combined into a single quantity, the
  ``bulge-statistic'' $\FGM = -0.693\mathrm{M}_{20} + 4.95\mathrm{G}
  -3.85$, where $F>0$ ($F<0$) indicates a bulge-dominated
  (disk-dominated) morphology. For details, consult
  \citet{Snyder2015b}. The galaxies in our quiescent sample are nearly
  all bulge-dominated, and therefore can be compared to
  observed early-type galaxies.}

We compare the sizes of Illustris galaxies to the observations from
the Millennium Galaxy Catalogue \citep{Poggianti2013} and the Sloan
Digital Sky Survey \citep{Dutton2011} in \changed{panel (c)} of Figure
\ref{fig.sample}. The sizes of low-mass ($\Mstar \lsim
10^{10.7}\Msun$) Illustris galaxies are known to be somewhat larger
than expected from observations \citep{Snyder2015b}. At the high-mass
end ($\Mstar \gsim 10^{12}\Msun$), the galaxies also appear to be
slightly too large, an effect that we believe is due to Intracluster
Light (ICL). Separation of the ICL from central galaxies is still an
open problem without a clear solution or commonly-adopted approach
\citep{Bernardi2013}, which complicates comparisons between models and
observations. Many ICL stellar particles are included as part of the
most massive central galaxies by {\sc Subfind} and, we attempt no
removal of this component, which contributes to the inflated sizes
($\Reff > 30\;\mathrm{kpc}$). In our subsequent analysis, we account
for the size discrepancies described here by considering ranges
defined by effective radii rather than physical units. Assuming that
Illustris galaxies are inflated in an approximately self-similar way
-- i.e.~scaled up by a constant factor, but otherwise matching the
structure of observed galaxies -- this should prevent the large sizes
from biasing our results. \changed{Inclusion or removal of the handful
  of massive galaxies with inflated sizes ($\Reff > 30\;\mathrm{kpc}$)
  do not affect the results that follow, so we choose to include them in the
  analysis.}

\changed{We compute the mean $\textrm{g-r}$ colors, as well as V-band weighted
  mean stellar metallicity ($\log\;(Z/Z_\odot)$) and stellar age of
  the Illustris galaxies, all measured within
  $\frac{1}{2}\;\Reff$. These properties are shown in panels (d), (e),
  and (f), respectively, of Figure \ref{fig.sample}. The quiescent
  galaxies have red colors, as is to be expected from their low
  star-formation rates.} They match reasonably the observed
mass-metallicity and mass-age relations. The most massive Illustris
galaxies appear slightly too metal-poor, which may be due to the
influence of the ICL. The inflated radii could bias low the
metallicities measured within $\frac{1}{2}\Reff$ because metallicity
decreases with radius (see following section).

%----------------------
\subsection{Measuring observable gradients}
\label{s.Methods.gradients}
In this paper, we consider the radial profiles of metallicity
($\mathrm{[Z/H]}$), stellar age, V-band surface brightness
($\Sigma_V,\; \mathrm{L}_\odot\,\mathrm{kpc}^{-2}$), and $\text{g-r}$
colors in our sample of quiescent Illustris galaxies. The V-band
surface brightness and $\textrm{g-r}$ colors are computed via the
stellar population synthesis model of \citep{Bruzual2003}, and assume
no dust extinction. The logarithmic gradients we compute are defined
as follows:
\begin{align}
  \mgrad &= \frac{d [\mathrm{Z}/\mathrm{H}]}{d \log R}\\
  \agrad &= \frac{d \mathrm{age}}{d \log R}\\
  \lumgrad &= \frac{d \log \Sigma_V}{d \log R}\\
  \cgrad &= \frac{d (\textrm{g-r})}{d\log R}.
\end{align}
A negative $\mgrad$, $\agrad$, $\lumgrad$, or $\cgrad$ therefore
represents outskirts that are more metal-poor, younger, less bright,
or bluer respectively than the interior.

These stellar population gradients are measured within three radius
ranges, which we define as: \textit{inner galaxy} ($0.1-1\;\Reff$),
\textit{outer galaxy} ($1-2\;\Reff$), and \textit{stellar halo}
($2-4\;\Reff$). \changed{We note that in \S\ref{s.grads_hist} we also
  discuss gradients computed farther ($2-10\;\Reff$) into the stellar
  halo, and find that the information content of the gradients is
  maximized at large radii. However, we concentrate our analysis on
  regions within $4\Reff$ where future observations will be most able
  to probe.}

Within each radius range, we compute the stellar properties in 5
logarithmic bins of radius, and measure a gradient using least-squares
fitting. This approach locally models the profiles as power-laws
within the chosen radial ranges, and thus the resulting gradients can
be considered the average slopes of the logarithmic profiles over the
given ranges.

\changed{In \S\ref{s.grads_hist}, we show that $\cgrad$ is strongly dependent
on the combination of $\mgrad$ and $\agrad$ and that, due to weak
correlations between the latter two, there is very little variation in
$\cgrad$ in the stellar halo. Therefore, we do not analyze $\cgrad$
further.}

In order to fairly compare to observations, we incorporate two
significant observational effects. All quantities are
luminosity-weighted using the V-band magnitude of each star particle,
which has significant effects on measured ages. We additionally
include projection effects by projecting the positions of all star
particles against a random line-of-sight. Ignoring this can introduce
significant biases between simulated measurements and
observations. For example, geometrical effects cause the projected
effective radius $\Reff$ of Illustris galaxies to be biased smaller than the 3D radius by
around $25\%$, which matches predictions from the
\citet{Hernquist1990} profile. We discuss the influence of projection
effects, and the uncertainties introduced into measurements of stellar
population gradients from random line-of-sight projections, in
Appendix \ref{a.projection}.

\begin{figure*}
  \plotoneman{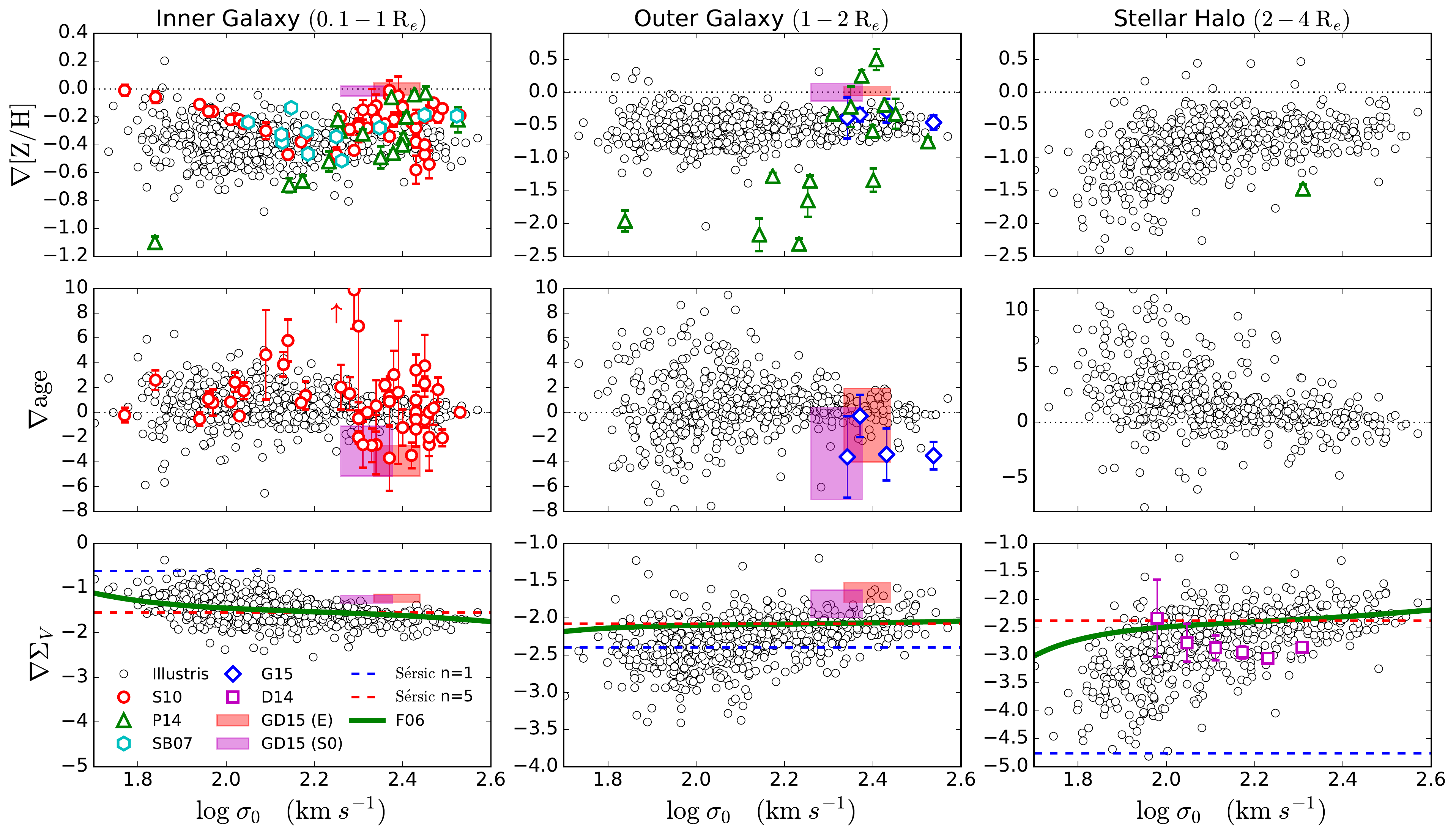}{0.98}
  \caption{Measured stellar-population gradients as a function of
    central velocity dispersion $\sigma_0$, with columns showing three
    radius ranges. Dotted black lines show a flat
    gradient. Observational comparisons: S10 \citep{Spolaor2010b}; P14
    \citep{Pastorello2014}; SB07 \citep{Sanchez-Blazquez2007}; G15
    \citep{Greene2015}; D14 \citep{DSouza2014}; GD15
    \citep{GonzalezDelgado2015}; F06
    \citep{Ferrarese2006}. \textit{Top:} Metallicity gradients are
    almost all negative with no significant mass-dependence, except in
    the stellar halo where low-mass galaxies have steeper
    gradients. Our measurements in the outer galaxy do not agree with
    the observations of \citet{Pastorello2014}; see Appendix
    \ref{a.SLUGGS} for discussion. \textit{Middle:} Age gradients are
    roughly flat, except in the halo where they tend to be
    positive. \textit{Bottom:} Surface-brightness profiles in low-mass
    galaxies steepen significantly from the inner galaxy to the
    stellar halo, while more massive galaxies have relatively constant
    gradients at all radii. Example gradients for two S\'ersic
    profiles ($n=1$ and $n=5$) are shown (dashed blue and red lines,
    respectively), and we also include comparisons to the S\'ersic
    index fits of \citep{Ferrarese2006} (green curve).}
  \label{fig.obs}
\end{figure*}

%----------------------------------
\subsection{Quantifying accretion histories}
\label{s.Methods.histories}
We quantify the accretion histories of the quiescent Illustris
galaxies in two ways: the \textit{ex-situ fraction} ($\fex$) and the
\textit{mean merger mass-ratio} ($\mu$), both measured within the
three radius ranges defined above.

Stars in each system are tagged either as \textit{in-situ} (formed
within their current host galaxy) or \textit{ex-situ} (formed in a
galaxy that subsequently merged with or was stripped by the current
phost). More details on this particle classification scheme can be
found in \citet{Rodriguez-Gomez2016}. The local ex-situ fraction (also
called the \textit{accreted fraction}) is thus the proportion of
stellar mass in a given region of a galaxy that was accreted from
smaller objects. It characterizes the total influence of all mergers
on the evolution of a particular galaxy region.

For each star tagged as ex-situ, we calculate the merger mass-ratio
between the host galaxy and the galaxy that brought in the star,
measured at the time when the latter reached its maximum stellar mass
\citep[see ][]{Rodriguez-Gomez2015a}. $\mu$ is the average of this
ratio over all stars in each of the three radius ranges, and thus
characterizes the relative influence of major and minor mergers to the
accretion history.

%======================================================
\section{Results}
\label{s.Results}
%------------------------
\subsection{Stellar population gradients}
\label{s.compare}
We begin with a comparison of the gradients of stellar populations
measured in Illustris quiescent galaxies to observations. Figure
\ref{fig.obs} shows the gradients in the quiescent galaxies as a
function of velocity dispersion within $\frac{1}{8}\Reff$ ($\sigma_0$,
to compare with observations) in the three radius ranges defined
above, with comparisons to available observations.

\begin{figure*}
  \plotoneman{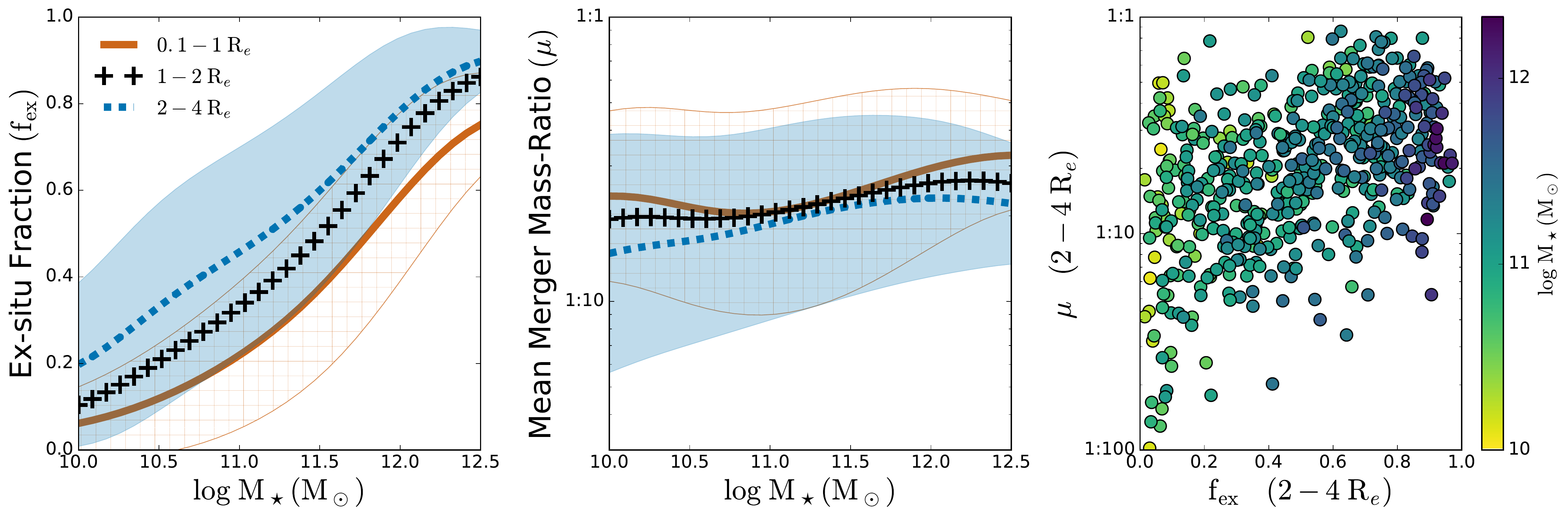}{0.95}
  \caption{Summary of the accretion history properties of the
    quiescent Illustris galaxy sample. \textit{Left:} The distribution
    of local ex-situ mass fractions in the quiescent sample, as a
    function of mass and in three radius ranges: inner galaxy ($0.1 -
    1 \;\Reff$, mean: orange solid line, std: vertical/horizontal
    hatches), outer galaxy ($1 - 2 \;\Reff$, mean: black crosses, std:
    not shown), and stellar halo ($2 - 4 \;\Reff$, mean: blue dashed
    line, std: solid blue). The distributions are smoothed
    using a Gaussian kernel with width $\sigma=0.3$ dex. The typical
    ex-situ fraction at any radius increases monotonically with mass
    (a clear sign of hierarchical growth) but there is a wide variety
    of individual accretion histories. At all masses, accreted
    material contributes more to the outskirts than the galactic
    interiors. \textit{Center:} The mass-weighted, mean merger
    mass-ratio, computed over the three radius ranges, as a function
    of stellar mass. Colors and smoothing are as in the left
    plot. Higher values represent accretion histories more influenced
    by major mergers. Galaxies show a wide variety of mean merger
    mass-ratios at all masses, but minor mergers tend to
    preferentially impact the largest radii. \textit{Right:} The
    relation between the ex-situ fraction and the mean merger
    mass-ratio in the stellar halo, with points color-coded by stellar
    mass. Galaxies that have accreted more of their stellar halo mass
    ($\fex \approx 1$) tend to have somewhat larger mean merger
    mass-ratios, implying it is uncommon to accrete large amounts of mass
    via minor mergers alone.}
  \label{fig.Accretion}
\end{figure*}

The top panels of Figure \ref{fig.obs} show metallicity
gradients ($\mgrad$). Almost all Illustris metallicity gradients are negative, in
agreement with observed early-type galaxies
\citeeg{Sanchez-Blazquez2007, Pastorello2014, GonzalezDelgado2015,
  Greene2015}. In the inner galaxy region, we find no significant
mass-dependence, unlike \citet{Spolaor2010b}, who found evidence of a
tight correlation between $\mgrad$ and $\sigma_0$ at low
masses. Likewise, we find no strong mass-dependence in the outer
galaxy region. Our measurements match the stacked observations from
the MASSIVE Survey \citep{Greene2015} at large masses, but we do not
find steep metallicity gradients ($\nabla\mathrm{[Z/H]} < -1.5$)
at low masses like those measured by the SLUGGS survey
\citep{Pastorello2014}. However, the apparent discrepancies between the
gradients are not supported by the individual metallicity data points
(see Appendix \ref{a.SLUGGS}). In the stellar halo, there is a
noticeable trend with mass, in the sense that low-mass galaxies have
somewhat steeper metallicity gradients. Comparable observations are
scant.

The middle panels of Figure \ref{fig.obs} show gradients in stellar
ages ($\agrad$). Illustris galaxies have a wide variety of both negative and
positive age gradients. In the inner and outer galaxy regions, our
measurements are in rough agreement with available observations
\citep{Spolaor2010b, Greene2015, GonzalezDelgado2015}. There is a
modest tension at large masses in the outer galaxy region, where we do
not find predominantly negative age gradients, but so far this is within the
uncertainties of observations. In the stellar halo, age gradients tend
to be positive, but as with metallicity gradients, observations have
yet to reach these large radii.

\begin{figure*}
  \plotoneman{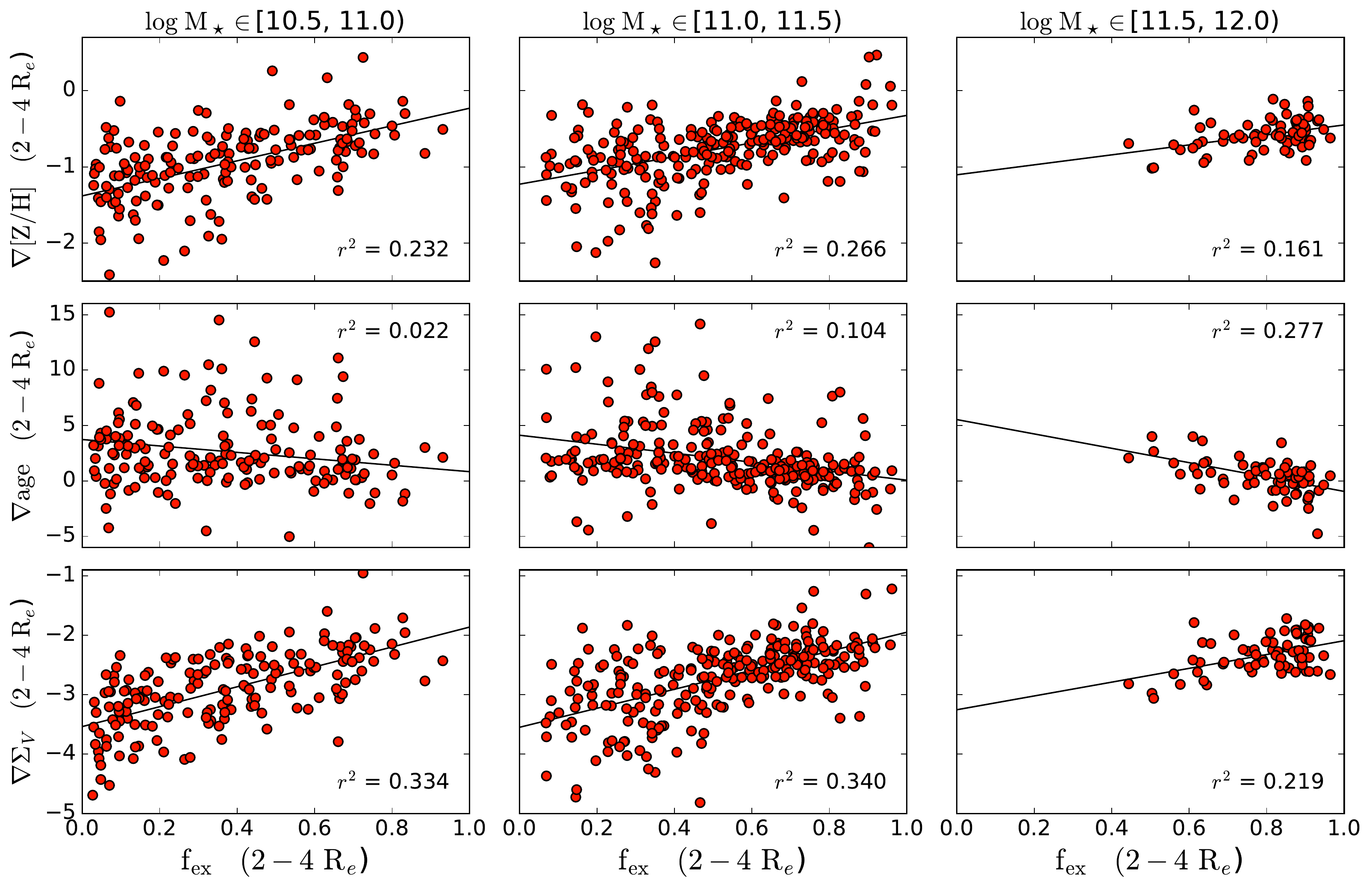}{0.90}
  \caption{Relations between stellar population gradients and local
    ex-situ fractions in the stellar halo ($2 - 4\;\Reff$). We select
    galaxies in three bins of mass (increasing from \textit{left} to
    \textit{right}), and show stellar halo metallicity (\textit{top}),
    age (\textit{middle}), and surface-brightness (\textit{bottom})
    gradients. The lines show the best-fit relation between the
    properties, denoted with the corresponding Pearson $r^2$. In all
    mass bins, there is a significant correlation between both the
    metallicity and surface-brightness gradients and the local ex-situ
    fraction. At fixed mass, stellar halos with larger accreted
    fractions tend to have flatter gradients ($\nabla\mathrm{[Z/H]}$
    or $\nabla\Sigma_V$ closer to $0$). There is at most a weak trend
    with $\nabla\mathrm{age}$, which exists only at large masses.}
  \label{fig.grads_v_frac_bins}
\end{figure*}

The gradients in surface-brightness profiles ($\lumgrad$) are shown in the bottom
panels of Figure \ref{fig.obs}. At all radii, surface-brightness
profiles are decreasing. We show comparisons to S\'{e}rsic profiles,
with average gradients computed in an similar fashion to our
measurements. Low-mass galaxies have relatively shallow
surface-brightness profiles in the inner galaxy region that become
successively steeper at larger radius, behavior that corresponds to
lower ($n\approx1.5$) S\'{e}rsic indices. In contrast, high-mass
galaxies have steeper profiles in the inner galaxy but show relatively
little steepening at large radii, in agreement with larger
($n\approx5$) S\'{e}rsic indices. Forthcoming results from MANGA,
SAMI, and CALIFA will soon improve observational constraints for the
inner and outer galaxy regions. In the stellar halo, stacked
observations of SDSS galaxies \citep{DSouza2014} appear somewhat
steeper than the Illustris sample, and have the opposite
mass-dependence to the simulations: slightly flatter profiles in the
lowest-mass bin. \citet{Ferrarese2006} found that the S\'{e}rsic index
increases with mass, in rough agreement with the trends we observe
here. However, Illustris stellar halo surface-brightness gradients at
low-mass are steeper than would be predicted from extrapolations of
the observed S\'{e}rsic profiles.

%------------------------
\subsection{Accretion histories}

The left panel of Figure \ref{fig.Accretion} shows the local ex-situ fraction ($\fex$)
in the three regions (inner galaxy, outer galaxy, and stellar halo) as
a function of mass. As is expected, accreted material makes up a
larger fraction of the total at large radii. Over all radius ranges,
the ex-situ fraction increases monotonically with mass. Comparisons
with lower-resolution runs of Illustris confirm the mass-dependence of
the ex-situ fraction is not due to resolution effects
\citep{Rodriguez-Gomez2016}, so this is a clear signature of
hierarchical growth.

The middle panel of Figure \ref{fig.Accretion} shows the wide
distribution of mean merger mass-ratios among galaxies of all masses
in Illustris. There is no significant trend with mass, but there is a
slight radial dependence, with the inner galaxy tending to be more
heavily influenced by major mergers than the outer regions. This
result is discussed in more detail in \citet{Rodriguez-Gomez2016}.

We also investigate the connection between the ex-situ fraction and
the merger mass-ratio in the stellar halo, which is shown in the right
panel of Figure \ref{fig.Accretion}. The overall patterns of
hierarchical accretion imply it is uncommon to accrete large amounts
of mass from minor mergers alone \citep{Rodriguez-Gomez2016}, which is
apparent in the rough correlation between $\fex$ and $\mu$ in the
stellar halo. Yet the large sample size of Illustris allows us to
study a small number of these uncommon galaxies (right panel of Figure
\ref{fig.Accretion}, lower-right of diagram).

%------------------------
\subsection{Connecting gradients to accretion histories}
\label{s.grads_hist}
\begin{figure*}
  \plotoneman{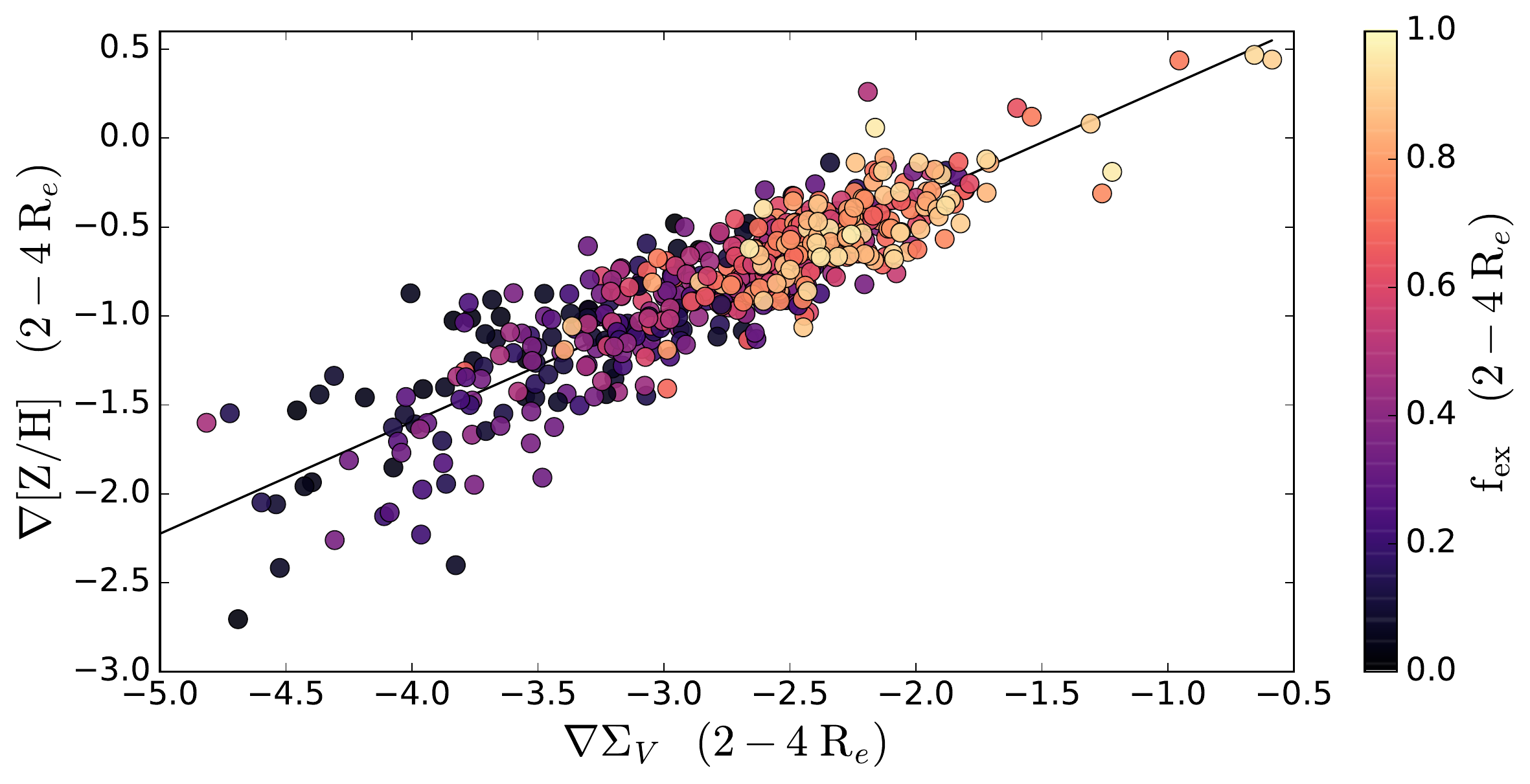}{0.90}
  \caption{Relation between metallicity and surface-brightness
    gradients in the stellar halo. The fit explains $\approx80\%$ of
    the variance in the data. The best fit
    (Eq.~\ref{eq.grad_correlation}) is shown as a black line. Data
    points are colored by the stellar halo ex-situ fraction, which
    largely determines a galaxy's location on the diagram. Galaxies
    with higher ex-situ fractions have both flatter metallicity and
    surface-brightness profiles.}
  \label{fig.grad_correlation}
\end{figure*}

We investigate how the range of stellar population gradients,
particularly in the stellar halo, is related to the variety of
galactic accretion histories. Figure \ref{fig.grads_v_frac_bins} shows
the trends between mass, stellar halo gradients, and stellar halo
ex-situ fraction. At fixed mass, the accreted fraction varies
significantly and is correlated with both $\nabla\mathrm{[Z/H]}$ and
$\nabla\Sigma_V$, as shown by Pearson $r^2$ coefficients in each panel
of Figure \ref{fig.grads_v_frac_bins}. Galaxies that have accreted
more of their halo stars have flatter metallicity profiles than those
that were dominated by in-situ growth. The same is true in regards to
the surface-brightness profiles, in agreement with previous results
\citep[Figure 5, lower-right panel of][]{Pillepich2014}.

Given that both the stellar halo metallicity and surface-brightness
gradients are correlated with the accreted fraction, we should expect
a tight relation between the two. In Figure
\ref{fig.grad_correlation}, we show this is indeed the case. In the
range $2 - 4\;\Reff$ the best fit line is given by:
\begin{equation}
  \nabla [\textrm{Z}/\textrm{H}] = 0.92 + 0.63 \, \nabla
  \Sigma_V \;.\label{eq.grad_correlation}
\end{equation}
The location of a galaxy along this relation depends on the amount of
accretion it has experienced in its stellar halo, as shown by the
colors in Figure \ref{fig.grad_correlation}. The fit explains $80.3\%$
of the variation in the data, as determined via the explained variance
regression score \citep{Wall2012}.

\begin{figure}
  \plotoneman{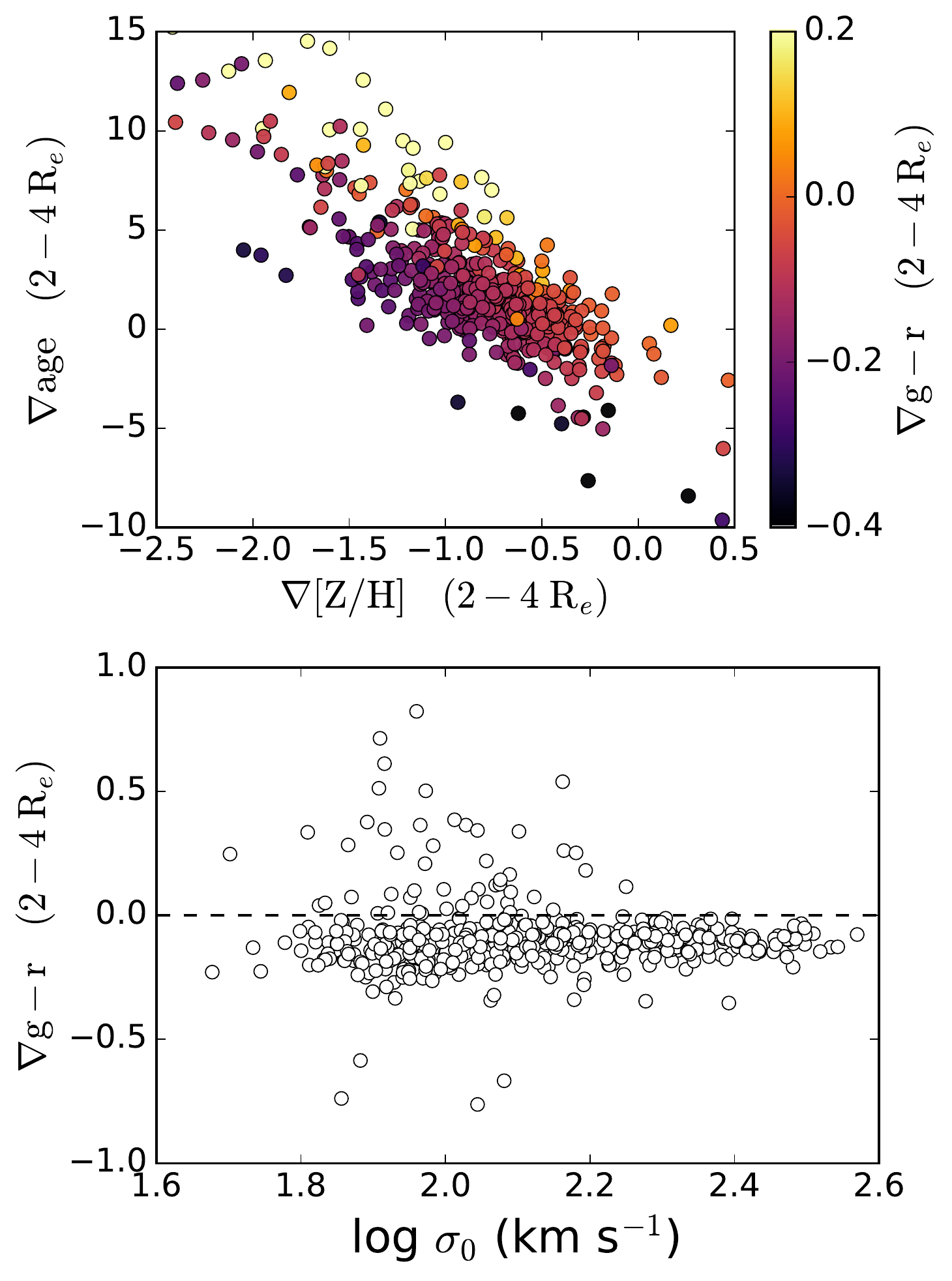}{0.98}
  \caption{\changed{Measured gradients in $\textrm{g-r}$ color in the
      stellar halo ($2 - 4 \;\Reff$). \textit{Top}: Stellar halo age
      gradients versus metallicity gradients, with points colored by
      $\cgrad$. Stellar halos with positive age gradients also tend to
      have particularly steep (negative) metallicity gradients. This
      has the net effect of washing out any significant variation in
      $\cgrad$ (the large cloud of points with similar
      colors). \textit{Bottom}: Stellar halo color gradients as a
      function of central velocity dispersion. Due to the canceling effects of
      age and metallicity gradients, there is no trend with velocity dispersion and
      very little overall variation in color gradients; nearly all
      values fall around $\cgrad \approx
      -0.1\;\mathrm{mag}\,\mathrm{dex}^{-1}$.}}
  \label{fig.colors}
\end{figure}

\begin{figure}
  \plotoneman{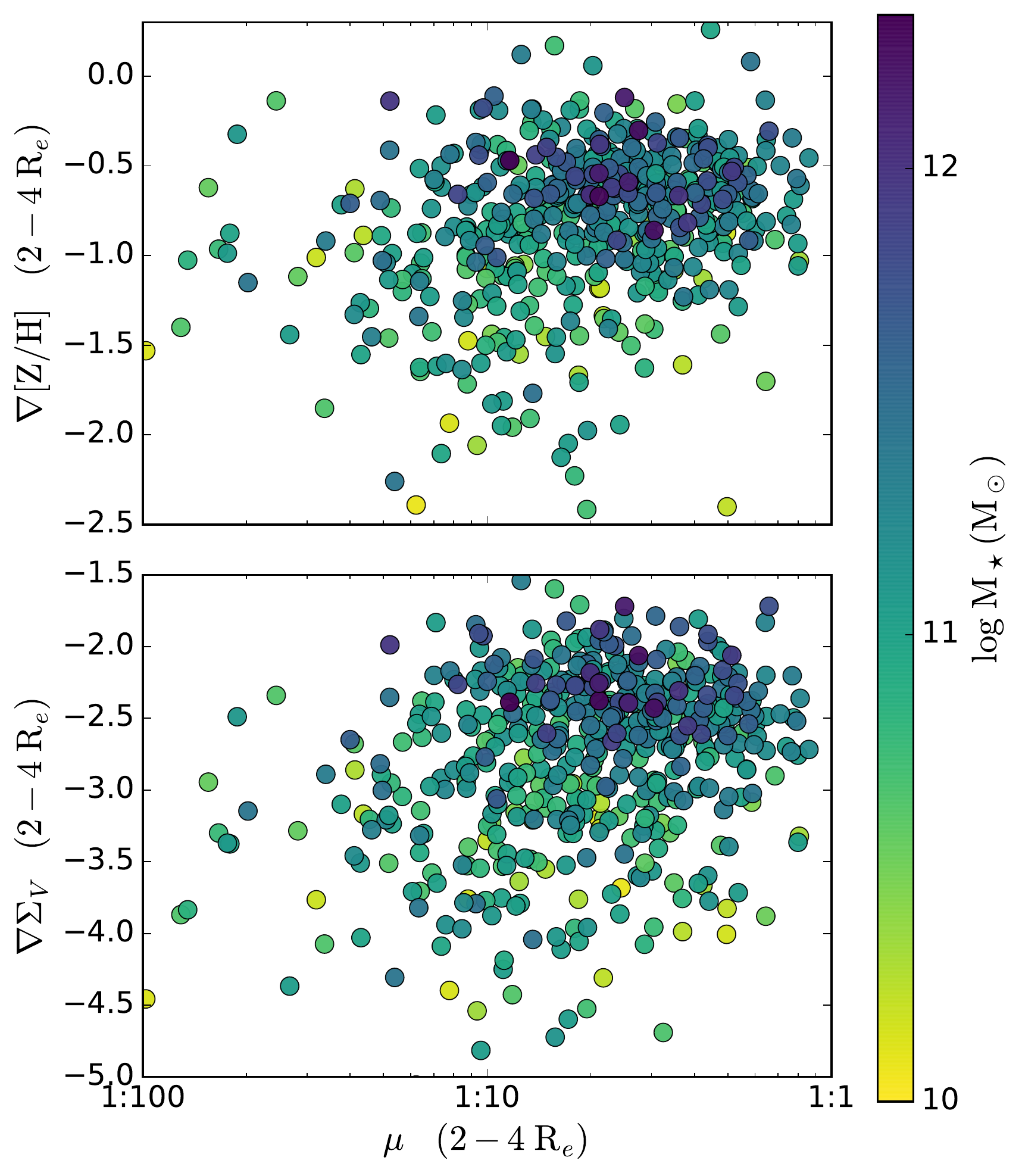}{0.98}
  \caption{Metallicity (\textit{top}) and surface-brightness
    (\textit{bottom}) gradients in the stellar halo plotted against
    the mean merger mass-ratio in the same region. Points are color
    coded by the galaxy stellar mass. The lack of significant
    correlation between either gradient and the mean merger mass-ratio
    (compare with Figure \ref{fig.grads_v_frac_bins}) implies that the
    gradients are not significantly influenced by whether accretion
    came in the form of major or minor mergers.}
  \label{fig.histories}
\end{figure}

The tight correlation between $\mgrad$ and $\lumgrad$ in the stellar
halo implies that the information content between the two is
similar. Both profiles are tracing the accretion history of the
stellar halo: halos with higher accreted fractions have both flatter
metallicity and surface-brightness profiles.

\changed{The weak correlation between $\agrad$ and $\fex$ suggests
  that the age profile retains relatively little information
  content. There is still, however, a slight correlation between
  $\agrad$ and $\mgrad$ in the stellar halo, as shown in the top panel
  of Figure \ref{fig.colors}. Importantly, the direction of this
  correlation results in a tendency to wash out any variation
  gradients of $g-r$ color ($\cgrad$) in the stellar halo. As shown in
  the bottom panel of Figure \ref{fig.colors}, nearly all the stellar
  halos in our quiescent sample have identical color gradients:
  $\cgrad \approx -0.1\;\mathrm{mag}\,\mathrm{dex}^{-1}$. For this
  reason, we do not consider color gradients further as a probe of
  accretion history properties.}

\begin{figure*}
  \plotoneman{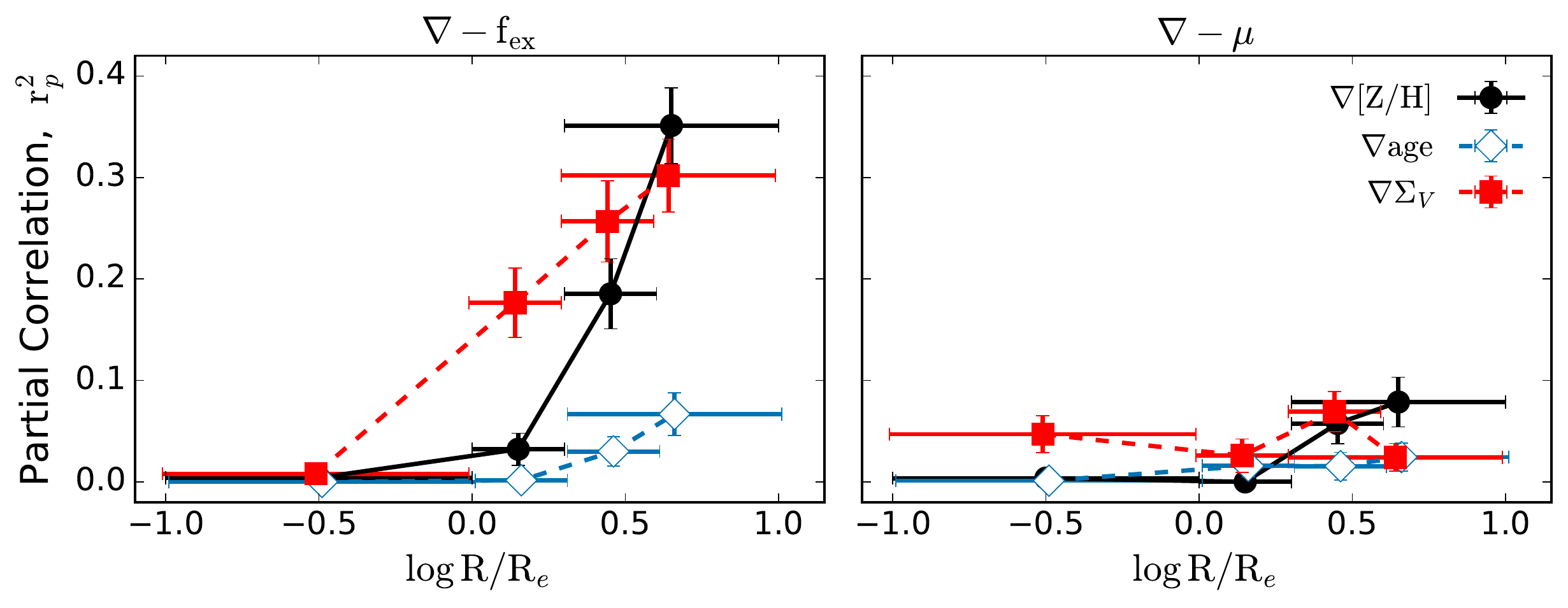}{0.95}
  \caption{Partial correlation coefficient ($r_p^2$, see text) between
    stellar population gradients ($\nabla$) and the accretion
    properties ($\fex$, \textit{left}, and $\mu$, \textit{right})
    computed as a function of radius range. The partial correlation
    coefficient represents the information content retained in each
    gradient (see text), and in each radius range is calculated
    between $\nabla\mathrm{[Z/H]}$ (black circles, solid line),
    $\nabla\mathrm{age}$ (open blue diamonds, dashed line), or
    $\nabla\Sigma_V$ (red squares, dashed line) and the accretion
    properties. The vertical error-bars are computed via
    bootstrapping. \textit{Left:} In the inner galaxy range ($0.1 - 1
    \mathrm{R}_e$), the gradients are not correlated with ex-situ
    fraction at all. The correlation increases with radius, from the
    outer galaxy ($1 - 2 \Reff$) to the stellar halo ($2 - 4 \Reff$),
    and is strongest when measured from $2 - 10 \Reff$, although
    $\nabla\mathrm{age}$ is never strongly correlated with the local
    ex-situ fraction. \textit{Right:} None of the stellar population
    gradients are good probes of the mean merger mass-ratio at any
    radius.}
  \label{fig.r2}
\end{figure*}

\changed{In addition to the cumulative amount of accretion, we also}
investigate the influence of merger mass-ratio on the stellar
population gradients. In Figure \ref{fig.histories} we show the
correlations between the mean merger mass-ratio and metallicity and
surface-brightness gradients, with all quantities measured in the
stellar halo. While both gradients are correlated with $\fex$, neither
is significantly correlated with the mean merger mass-ratio. Stellar
halos with a given stellar population gradient are just as likely to
have been dominated by major mergers as by minor mergers.

To quantify the amount of information retained in each of the stellar
population profiles, we compute the first-order partial correlation
coefficients $r_p$ \citep[\S4.3,][]{Wall2012} between the observable
gradients and the accretion properties. The partial correlation
coefficient between variables $x_1$ and $x_2$, controlling for the
influence of $x_3$, is given by:
\begin{equation}
  r_p(x_1, x_2) = \frac{r_{12} - r_{13}r_{23}}{\sqrt{(1 - r_{13}^2)(1
      - r_{23}^2)}}\,,
\end{equation}
where $r$ is the Pearson correlation coefficient. Thus, $r_p$
represents the correlation between the two variables beyond what can
be explained from mutual correlation with the third
(potentially-confounding) variable.

In Figure \ref{fig.r2}, we show $r_p^2$ as a function of galactic
radius, computed between the three gradients ($\nabla$) and the local
accretion history parameters ($\fex$ and $\mu$). We control for the
confounding influence of stellar mass ($\log\mathrm{M}_\star$), which is
shown in Figure \ref{fig.Accretion} to be correlated with the ex-situ
fraction. The uncertainty is quantified using $10,000$ bootstrap
resamplings. Therefore, $r_p^2$ represents the amount of variation in
each gradient that is solely explained by the accretion properties.

Comparing the left and right panels of Figure \ref{fig.r2}, the
gradients correlate much more strongly with the ex-situ fraction than
with the mean merger mass-ratio. It is apparent that the stellar
population gradients are tracing the total amount of accretion, not
whether that accretion was via major or minor mergers.

Figure \ref{fig.r2} also shows that the information content of
accretion is retained in the stellar population profiles only at very
large radii from the galaxy. In the inner galaxy region ($0.1 -
1\;\Reff$), there is no correlation between the accreted fraction and
any of the gradients. At successively larger radii,
$\nabla\mathrm{[Z/H]}$ and $\nabla\Sigma_V$ become increasingly
correlated with the local ex-situ fraction, with $\fex$ explaining as
much as $30-40\%$ of the variation in the gradients when measured very
far ($2 - 10 \;\Reff$) into the stellar halo. The metallicity gradient
$\mgrad$ does not retain more information content than the
surface-brightness profile $\lumgrad$, except possibly in the very
farthest regions of the stellar halo. $\nabla\mathrm{age}$ is not a
strong probe of the accretion information content at any radius. These
findings suggest that deep photometry in galactic stellar halos should
be equal probes of galactic accretion histories as spectroscopic
metallicity measurements.

%------------------------
\subsection{Redshift evolution of stellar population profiles}
\label{s.Redshift}

\begin{figure*}
  \plotoneman{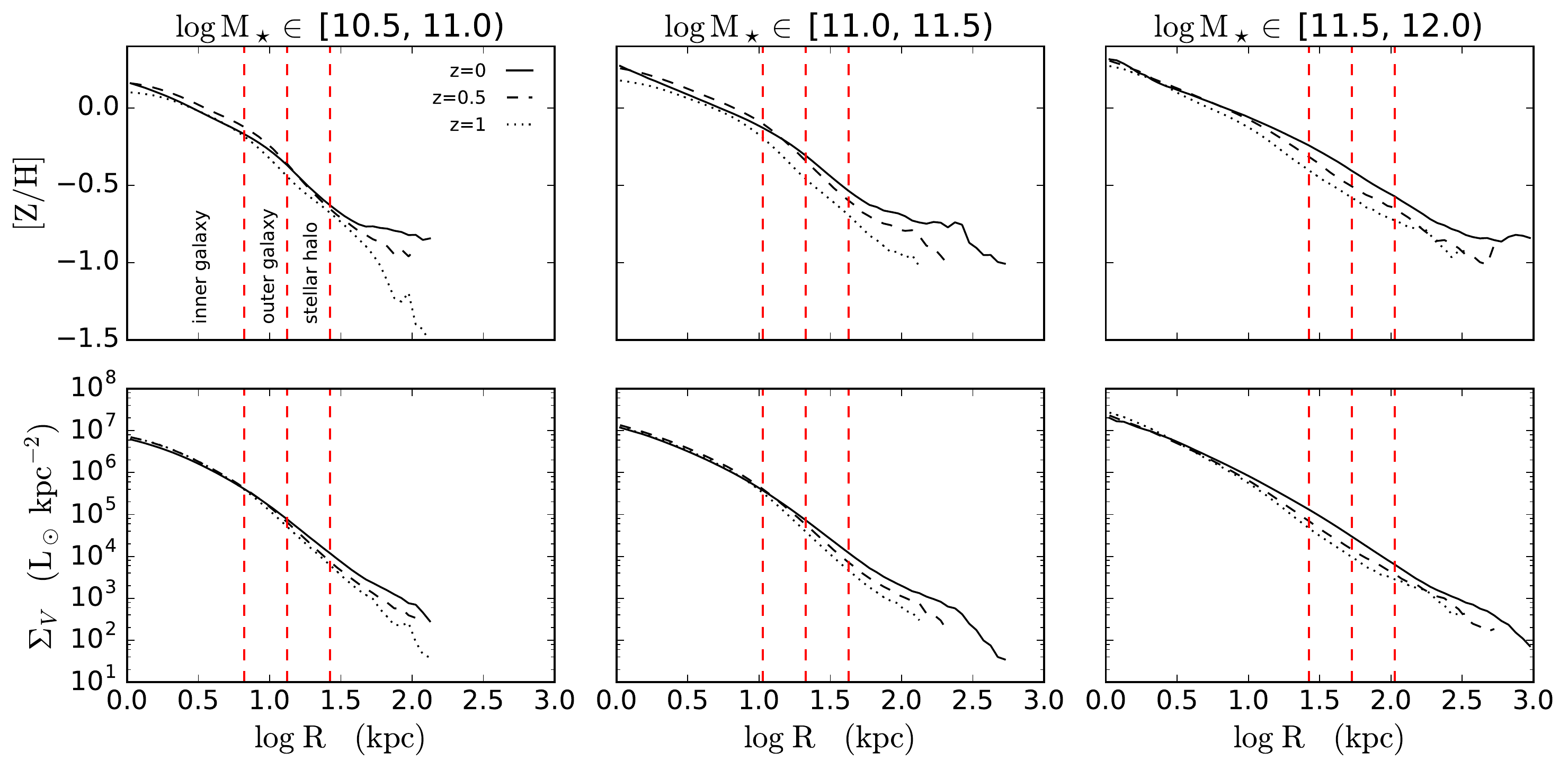}{0.95}
  \caption{Redshift-evolution of mean metallicity (\textit{top})
    and surface-brightness (\textit{bottom}) profiles. The projected
    profiles of $z=0$ galaxies (\textit{solid}) and their $z=0.5$
    (\textit{dashed}) and $z=1$ (\textit{dotted}) progenitors are
    stacked in three bins -- increasing from left to right --
    according to the $z=0$ mass. The red dashed lines show 1, 2, and 4
    times the mean $\Reff$ (measured at $z=0$) in each mass range,
    representing the typical boundaries of the inner galaxy, outer
    galaxy, and stellar halo regions. The central regions ($\mathrm{R}
    \lsim 10 \;\mathrm{kpc}$) of the profiles do not evolve
    significantly. In the outskirts, however, the mean profiles at all
    masses flatten over time, representing the accretion of ex-situ
    material in the stellar halo. Particularly at lower masses, there
    is even more significant change beyond $4\Reff$.}
  \label{fig.redshift_profiles}
\end{figure*}

We have shown that Illustris stellar halos with larger accreted
fractions tend to host flatter gradients in metallicity and
surface-brightness. If this is a causal relation -- i.e.~accretion
actively flattens these stellar population profiles -- then we would
expect most galaxies to have steeper profiles at earlier times, when
they had accreted less material. To investigate this, we study
the evolution of metallicity and surface-brightness profiles as a
function of redshift.

We use the Illustris SubLink merger trees
\citep{Rodriguez-Gomez2015a} to identify the $z=0.5$ and
$z=1$ progenitors of each galaxy in our quiescent sample. We project
the positions of particles in each progenitor galaxy against a random
line-of-sight, then average the profiles at each redshift in three
bins, according to the $z=0$ mass. The results are shown in Figure
\ref{fig.redshift_profiles}. We note that faithfully comparing the
evolution of these profiles to observations will be difficult because
of progenitor bias \citep{VanDokkum2001, Wellons2016a}.

\begin{figure}
  \plotoneman{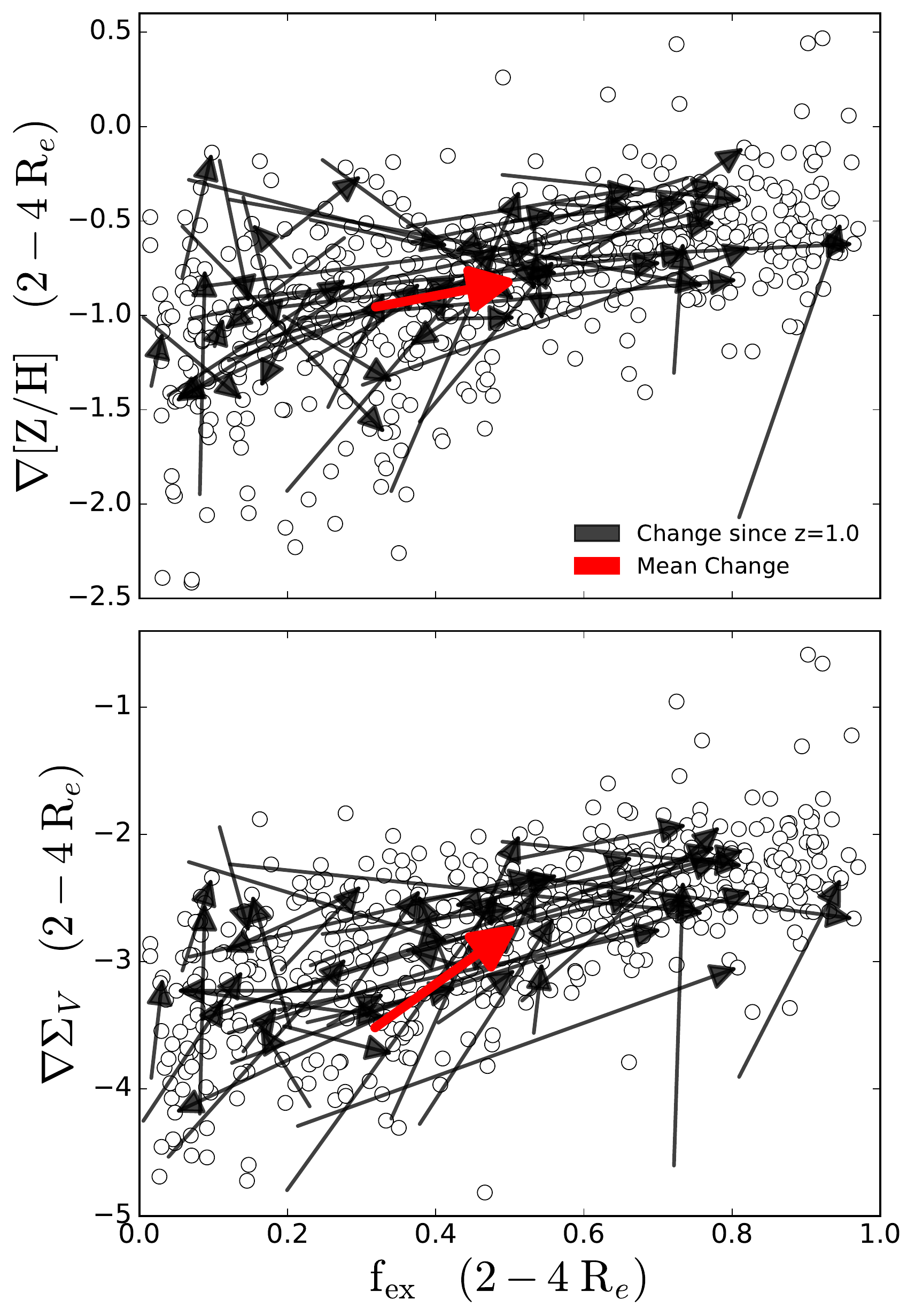}{0.98}
  \caption{Individual evolutionary paths of galaxies in terms of
    stellar halo gradients and local ex-situ fractions. The $z=0$
    relation between stellar halo gradient and ex-situ fraction is
    shown as white points. The evolutionary paths from $z=1$ for a
    random sample of 50 galaxies are shown as black arrows, with the
    mean change of the entire population in red. Galaxies have a wide
    variety of evolutionary histories, but on average they tend to
    evolve along the $\nabla - \mathrm{f}_\mathrm{ex}$ relation.}
  \label{fig.redshift_arrow}
\end{figure}

From $z=1$ to the present, none of the profiles evolve significantly
over the inner $\approx 10\;\mathrm{kpc}$. This is particularly
interesting in the case of the surface-brightness profile, as the
inner surface-\textit{density} profiles (\textit{not shown}) do
increase by a factor of $\approx 3$ over this time period. The
combination of accretion and passive evolution of an old stellar
population in these central regions appears to roughly cancel out any
changes to the inner $10\;\mathrm{kpc}$ of the surface-brightness
profiles.

In the outer regions of the galaxies, the metallicity and
surface-brightness profiles flatten noticeably since $z=1$. The
galaxies that are quiescent at $z=0$ were not necessarily quenched at
higher redshifts, so this does not only represent ex-situ mass
growth. However the lack of any significant change to the inner
profiles (where in-situ growth would show substantial effects)
suggests that the majority of this evolution comes from the accretion
of material into the galactic outskirts. In the two lower mass bins,
the profiles steepen even more significantly beyond $4\;\Reff$,
supporting the indications from Figure \ref{fig.r2}: observations that
push into the farthest reaches of the stellar halo will be the most
successful at constraining the history of accretion.

In Figure \ref{fig.redshift_arrow}, we show the individual
evolutionary paths since $z=1$ of a random sub-sample of our quiescent
Illustris galaxies, in terms of stellar population gradients and
ex-situ fraction in the stellar halo. Also shown is the $z=0$ relation
between the gradients ($\nabla$) and accreted fraction
($\mathrm{f}_\mathrm{ex}$). Galaxies have diverse histories, but on
average stellar halo gradients have gotten flatter while ex-situ
fractions have increased. The majority tend to move along the $\nabla
- \mathrm{f}_\mathrm{ex}$ relation observed at $z=0$.

%======================================================
\section{Discussion - The Information Content of Stellar Halos}
\label{s.Discussion}

We find that, at a fixed mass, galaxies with different average
accretion histories will have different stellar halo
properties. Steeper profiles from $2-4\;\Reff$ in either metallicity
or surface-brightness indicate a quieter history of accretion into the
halo, while relatively flat profiles signal that a stellar halo was
built-up from large amounts of accretion. \changed{As shown in Figure
\ref{fig.r2}, this becomes even more significant when measured far
($2-10\;\Reff$) into the stellar halo.} This agrees with predictions
of the influence of in-situ and ex-situ growth on stellar population
gradients \citep{Kobayashi2004,Oser2010}, and with previous analysis
of the stellar mass-density profiles in Illustris stellar halos
\citep{Pillepich2014}. In-situ star formation in early-type galaxies
initially produces steep metallicity gradients and surface-brightness
profiles beyond $\approx 2\;\Reff$. Mergers and accretion tend to
deposit stars in the outskirts of the galaxies, flattening their
metallicity gradients and surface-brightness profiles. Age gradients
are poor indicators of a galaxy's accretion history.

A surprising discovery was that none of the stellar population
profiles studied are correlated with the local merger mass-ratio. It
is easy to imagine that a stellar halo that accreted a large amount of
mass via many $1$:$100$ mergers would have different properties from
one that experienced a single $1$:$1$ merger. Previous analysis of
Illustris galaxies \citep{Rodriguez-Gomez2016} has shown that minor
mergers tend to deposit their ex-situ material at larger radii than do
major mergers, which would necessarily affect the resulting
surface-brightness profiles. \citet{Amorisco2016a} used a library of
N-body merger simulations to model the \textit{accreted} portions of
stellar halos, and found the slopes and normalizations of the ex-situ
mass profiles correlate with the typical merger mass-ratio.

The mass-metallicity relation for galaxies likewise suggests that
major mergers (involving massive galaxies) should typically leave
different signatures in metallicity profiles than minor mergers (with
low-mass satellites). Indeed, previous hydrodynamical zoom simulations
of 10 galaxies by \citet{Hirschmann2015} found that metallicity and
color gradients beyond $2\;\Reff$ were correlated with the mass gained
from major mergers (or alternatively the overall mean merger
mass-ratio), implying that major mergers tended to flatten
gradients more significantly than minor mergers.

Our findings instead suggest that the primary driver of observable
stellar halo metallicity and surface-brightness profiles is the total
amount of accreted mass, regardless of the typical merger
mass-ratio. When we separate the stellar halo stars into in-situ and
ex-situ populations, we find that the slope of the ex-situ
surface-brightness profile (not shown) \textit{does} depend on the
typical merger mass-ratio, in agreement with
\citet{Amorisco2016a}. However the in-situ surface-brightness profile
(which does not depend on merger mass-ratio) is significantly steeper
than the ex-situ profiles. Therefore, the ratio of ex-situ to in-situ
material is what primarily determines the slope of the overall
profile.

The correlation between metallicity gradient and merger mass-ratio
found by \citet{Hirschmann2015}, which is not apparent in Illustris
stellar halos, could be due to their relatively small sample size. As
shown in the right panel of Figure \ref{fig.Accretion}, stellar halos
with high ex-situ fractions generally tend to have been dominated by
major mergers. It is uncommon to accrete a significant amount of mass
from minor mergers alone. A re-examination of the
\citet{Hirschmann2015} data supports this conclusion: the galaxies in
their simulations with high merger mass-ratios had higher accreted
fractions overall. Thus the correlation they observed between $\mgrad$
and $\mu$ is actually due to the underlying correlation between $\mgrad$ and
$\fex$.

With the large sample size of Illustris we are able to offer improved
statistics, including a handful of the uncommon galaxies that accreted
significant fractions of their stellar halos from minor mergers alone
(right-most panel of Figure \ref{fig.Accretion}; points in lower-right
of diagram). These high $\fex$, low $\mu$ stellar halos have
relatively flat metallicity and surface-brightness profiles, which
supports the model that the ex-situ fraction is the dominant driver of
stellar halo profiles, not the merger mass-ratio.

There are known issues with the Illustris model, as discussed in
\S\ref{s.Intro}, which impact this work. The incomplete quenching of
star-formation at $z=0$ likely resulted in an under-population of the
red sequence, and thus a smaller sample of quiescent galaxies. Another
significant issue with Illustris is the inflated sizes of galaxies
below $\Mstar \lsim 10^{10.7}\Msun$. We have attempted to compensate for
this by considering regions defined by $\Reff$, rather than a fixed
range in $\mathrm{kpc}$. If Illustris galaxies retain the proper
relative structure, but are simply scaled large by a factor of a few,
this should be sufficient to cancel the systematic size bias. However,
it is possible that the inflated sizes (and correspondingly low
velocity dispersions) of small galaxies left them more susceptible to
tidal disruption, which could affect the amount of accreted stellar
material deposited into stellar halos.

These systematic issues in the Illustris galaxy formation model are
being addressed in a new generation of Illustris simulations, which
among other changes has new implementations of AGN and galactic wind
feedback \changed{\citep[][Pillepich et al. 2016 in prep]{Weinberger2016}} that are designed to improve the
efficiency of quenching. \changed{\citet{Hirschmann2015} (but also
  \citealt[][Figure 3]{Rodriguez-Gomez2016}) found that different
  galactic and AGN feedback prescriptions can have significant effects on the
  mass fraction of stellar accretion. However, while we expect such
  changes to directly affect the normalization of the brightness and
  metallicity profiles studied here, we argue that their impact on the
steepness of such profiles (the gradients, which are the focus of this
work) is less direct and possibly negligible. Analysis of these future simulations
will help identify possible biases to the quantitative results
presented here and to confirm whether our qualitative conclusions
about the information content of brightness and metallicity gradients
in the stellar halo hold.}

%======================================================
\section{Summary}
\label{s.Summary}
In this paper, we have investigated the observable stellar properties
of simulated early-type galaxies, and how stellar population gradients
are connected to the history of accretion in the stellar halo. We
considered a sample of quiescent galaxies from Illustris, a
state-of-the-art hydrodynamical cosmological simulation
\citep{Vogelsberger2014b, Genel2014a, Nelson2015a}. Our final sample
includes 537 quiescent galaxies ranging in stellar mass from $\Mstar =
10^{10} - 2\times10^{12} \Msun$, which we demonstrate have overall
ages and metallicities which agree with observations.

Developing an accurate understanding of how the stellar halo traces
accretion requires a statistical population of galaxies large enough
to sample the wide variety of possible accretion histories in a
$\Lambda$CDM cosmology. Illustris is thus an ideal laboratory to study
this problem. Its large cosmological volume avoids the issue of small
sample sizes found in smaller hydrodynamical simulations
\citeeg{Hirschmann2015}, while also producing realistic galaxies,
which are difficult to replicate using N-body only simulations
\citeeg{Cooper2010}. We now summarize our primary results:

\begin{enumerate}
\item
  Stellar population gradients in quiescent Illustris galaxies are in
  overall agreement with available observations\changed{, although there are
  few comparable observations at large radii ($>1\;\Reff$)}. We measure
  logarithmic gradients in metallicity ([Z/H]), age, and
  surface-brightness ($\Sigma_V$), as computed in three radius ranges:
  the inner galaxy ($0.1 - 1\;\Reff$), the outer galaxy
  ($1-2\;\Reff$), and the stellar halo ($2 - 4\;\Reff$). These
  measurements incorporate realistic observational effects and find
  overall negative metallicity and surface-brightness gradients as
  well as age gradients that are on-average flat.

\item
  At fixed mass, the gradients of both the metallicity and
  surface-brightness profiles beyond $2\;\Reff$ are strongly
  correlated with the overall amount of accretion (the ex-situ
  fraction) in the stellar halo, as well as with one-another. Age
  gradients are at most only weak tracers of accretion histories. The
  information content of accretion histories is only preserved in
  stellar population gradients in the stellar halo, beyond
  $\approx 2\;\Reff$.

\item
  Stellar halo metallicity profiles do not contain extra information
  related to accretion histories than do surface-brightness profiles,
  suggesting that photometric observations in stellar halos are
  sensitive to the same information content as spectroscopic
  measurements.
  
\item
  The total amount of accretion is what primarily affects profiles of
  stellar populations in the stellar halo, not the mean merger
  mass-ratio. In a $\Lambda$CDM cosmology, it is uncommon for a stellar
  halo to have accreted large amounts of mass from minor mergers
  alone. However, the few examples available in Illustris show relatively
  flat stellar halo metallicity and surface-brightness gradients
  corresponding to their high ex-situ fractions.

\item
  The only significant evolution of the metallicity and
  surface-brightness profiles since $z=1$ occurs at radii larger than
  $\approx 10\;\mathrm{kpc}$, where both profiles become flatter with
  time. While there is significant system-to-system variation, most
  stellar halos evolve towards higher accreted fractions and
  flatter stellar population profiles and tend to do so along the
  $\nabla - \mathrm{f}_\mathrm{ex}$ relation observed at $z=0$.
   
\end{enumerate}

Taken together, these findings show that the metallicity and
surface-brightness profiles in the stellar halo are shaped by the
cumulative history of accretion a halo has undergone. The initial
in-situ formation of a galaxy's stars leaves steep profiles, with few,
primarily metal-poor stars in the galactic outskirts. As the galaxies
grow through mergers, the accretion of material deposits significantly
more stars, which are relatively more metal-rich, into the stellar
halo, flattening $\mgrad$ and $\lumgrad$.

By calculating averaged radial profiles this work approximated
galaxies as having smooth radial distributions. Our results suggest
that these azimuthally-averaged profiles are not sensitive to the
relative influence of major and minor mergers. In practice, stellar
halos typically do not show azimuthal-symmetry, but instead have
significant clumpy substructure. In future analysis, we hope to
characterize this in Illustris stellar halos and determine whether
stellar halo substructure is a better probe of specific merger
histories than smoothed profiles.

Observational surveys designed to characterize the stellar halos of
individual galaxies can be approached in two basic ways. Deep
spectroscopic observations can measure metallicity and age gradients
\citeeg{Coccato2010,Pastorello2014,GonzalezDelgado2015}, while
integrated light photometry can constrain the surface-brightness
profiles to large radii \citeeg{Mihos2005,Duc2014,VanDokkum2014}. Our
results suggest that either approach is sufficient to infer the
overall amount of accretion that contributed to the build-up of a
galaxy's stellar halo. Future surveys which study a common sample of
stellar halos with both photometric and spectroscopic observations can
be used to test our prediction of a correlation between metallicity
and surface-brightness profiles (see Figure
\ref{fig.grad_correlation}). Through a combined effort, simulations
like Illustris and future observational surveys can help to decode the
information content preserved in stellar halos and solidify our
understanding of the formation and evolution of early-type galaxies.

\acknowledgments B.C.~wishes to thank the anonymous referee for
helpful, constructive comments and Daniel Eisenstein and Nelson
Caldwell for their input and recommendations in the early drafts of
this work, and acknowledges support from the NSF Graduate Research
Fellowship Program under grant DGE-1144152. C.C.~acknowledges support
from NASA grant NNX15AK14G, NSF grant AST-1313280, and the Packard
Foundation. L.H.~acknowledges support from NASA grant NNX12AC67G and
NSF grant AST-1312095.

\software{This research has made use of NASA's Astrophysics Data
  System Bibliographic Services, as well as the following software packages:
  Matplotlib \citep{Hunter2007}, IPython \citep{Perez2007}, SciPy
  \citep{Jones2001}, NumPy \citep{VanderWalt2011}, Pandas
  \citep{McKinney2010}, and Scikit-Learn \citep{Pedregosa2011}.}

\appendix
\section{Projection Effects}
\label{a.projection}

\begin{figure*}
  \plotoneman{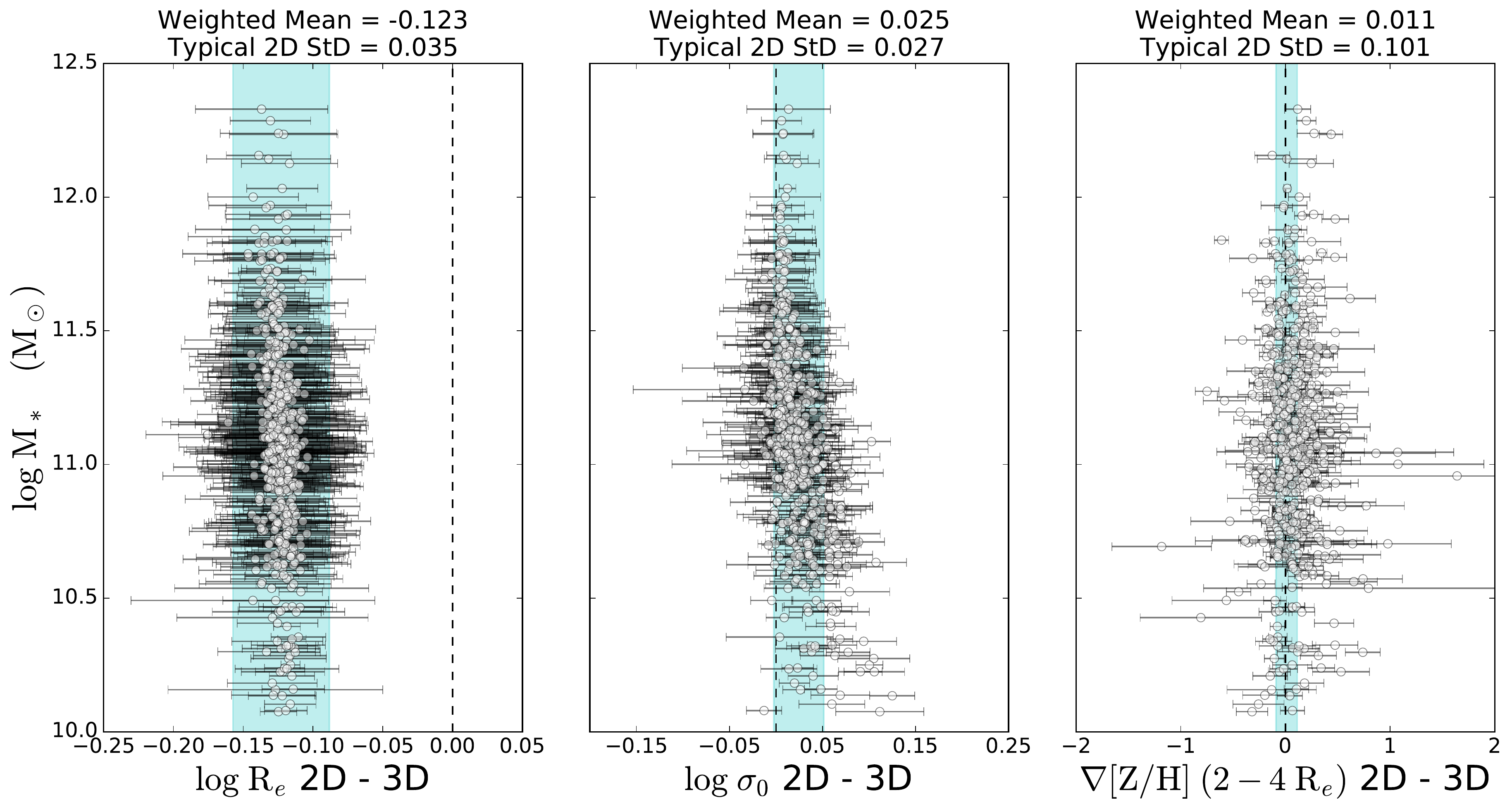}{0.99}
  \caption{The distribution of projection biases (mean of 100 random
    projections minus 3-D measurement) for a few measured quantities:
    galaxy sizes ($\log \Reff$, \textit{left}), central
    velocity dispersions ($\log\sigma_0$, \textit{middle}), and
    stellar halo metallicity gradients ($\nabla\mathrm{[Z/H]}$,
    \textit{right}). Horizontal error-bars represent the standard
    deviation from 100 random projections. The weighted-mean bias and
    the median 2D std (see text) are shown at the top, and are shown
    as cyan regions. Projection systematically biases galaxy sizes low
    by $\approx 25\%$, which is a well-understood geometrical
    effect. There is a small $< 0.1\;\mathrm{dex}$ systematic bias in
    velocity dispersion, with low-mass galaxies showing relatively
    larger biases. In the stellar halo, projection does not have a
    significant systematic bias on metallicity gradients, but there
    can be large variation ($>0.1\;\mathrm{dex}$) due to different
    line-of-sight projections.}
  \label{fig.projection}
\end{figure*}

All measurements in this paper are computed after projecting the
positions of the star particles in each galaxy against a random
line-of-sight. Ignoring this effect can introduce systematic biases
between simulations and observations. For example, simple geometrical
effects will cause the observed sizes ($\Reff$) of galaxies to be
smaller after projection than the 3D effective radii. Deviations from
spherical symmetry can also result in significant variation
depending on the particular line-of-sight projection, such as in the
case of significant ellipticity or a disk-like structure.

To study the impact of these effects, we compute the typical bias and
uncertainty that arises from comparing multiple projected measurements
to those computed in 3D. First, we compute $\Reff$ and other
subsequent properties ($\sigma_0$, gradients $\nabla$, local ex-situ
fractions $\mathrm{f}_\mathrm{ex}$ and mean merger mass-ratios $\mu$)
using the full 3D positions and velocities of all particles. We then
measure the same properties in projection 100 times, using different,
randomly selected lines-of-sight for each projection. Each iteration
is self-consistent: quantities such as $\nabla$,
$\mathrm{f}_\mathrm{ex}$, and $\mu$ are calculated in regions defined
by that iteration's $\Reff$. We then take the mean and standard
deviation of the 100 random projections. We study the difference
between the mean 2D value and the 3D value ($2D - 3D$), as well as the
median of the standard deviations between 100 2D projections ($2D
\;\mathrm{std}$). We do not compute the 3D gradient of the
surface-brightness profile $\lumgrad$, which is inherently a projected
quantity.

\begin{table}
  \begin{tabular}{|C|R|d{3.4}|d{0.3}|}\hline
  \textrm{Property} & \textrm{Region} & \multicolumn{1}{c|}{2D - 3D} &
  \multicolumn{1}{c|}{2D std}\\\hline \log \Reff &
  \multicolumn{1}{c|}{$-$} & -0.123 & 0.035\\\hline
  \log \sigma_0 & \leq \frac{1}{8}\;\Reff & 0.025 & 0.027\\\hline
  & 0.1 - 1 \;\Reff &  0.023 & 0.028\\
  \mgrad & 1 - 2\;\Reff & 0.043 &  0.067\\
  & 2 - 4 \;\Reff & 0.011 & 0.10\\\hline
  & 0.1 - 1 \;\Reff &  -0.058 & 0.25\\
  \agrad & 1 - 2\;\Reff & -0.058 &  0.49\\
  & 2 - 4 \;\Reff & 0.038 & 0.55\\\hline
  & 0.1 - 1 \;\Reff & \multicolumn{1}{c|}{$-$} & 0.065\\
  \lumgrad & 1 - 2 \;\Reff & \multicolumn{1}{c|}{$-$} & 0.099\\
  & 2 - 4 \;\Reff & \multicolumn{1}{c|}{$-$} & 0.13\\\hline  
  & 0.1 - 1 \;\Reff &  -0.002 & 0.004\\
  \mathrm{f}_\mathrm{ex} & 1 - 2\;\Reff & -0.0003 &  0.006\\
  & 2 - 4 \;\Reff & 0.004 & 0.01\\ \hline
  & 0.1 - 1 \;\Reff &  -0.0001 & 0.003\\
  \mu & 1 - 2 \;\Reff & -0.0012 & 0.004\\
  & 2 - 4  \;\Reff & -0.0008 & 0.005\\ \hline
  \end{tabular}
  \caption{The weighted-mean projection bias ($2D - 3D$) and median
    value of the 2D standard deviation (between the 100 random
    projections) for the measured properties in this paper. A negative
    bias means the projected value is less than the 3D value. Galactic
    effective radii ($\Reff$) are systematically biased low, by around
    $25\%$.  The average projection biases of all other measured
    quantities are small in magnitude ($<0.1\;\mathrm{dex}$), but
    there can be significant variation due to particular line-of-sight
    projections.}
  \label{tab.projection}
\end{table}

Figure \ref{fig.projection} shows the results of this analysis for a
few measured properties: galaxy sizes, velocity dispersions, and the
stellar halo metallicity gradients, and the results for all measured
quantities are listed in Table \ref{tab.projection}.

As predicted from geometrical arguments, projected galaxy sizes
($\log\Reff$) are biased low by about $25\%$ on average ($2D - 3D
\approx -0.12$), in excellent agreement with the analytical solution
for the \citet{Hernquist1990} profile. The projection biases in other
properties are relatively small ($2D - 3D <
0.1\;\mathrm{dex}$). However, different lines-of-sight projections can
induce significant variations ($2D\;\mathrm{std} > 0.1\;\mathrm{dex}$)
in measurements of stellar population gradients, especially in the
stellar halo.

\section{Comparison to SLUGGS Survey Gradients}
\label{a.SLUGGS}

\begin{figure*}
  \plotoneman{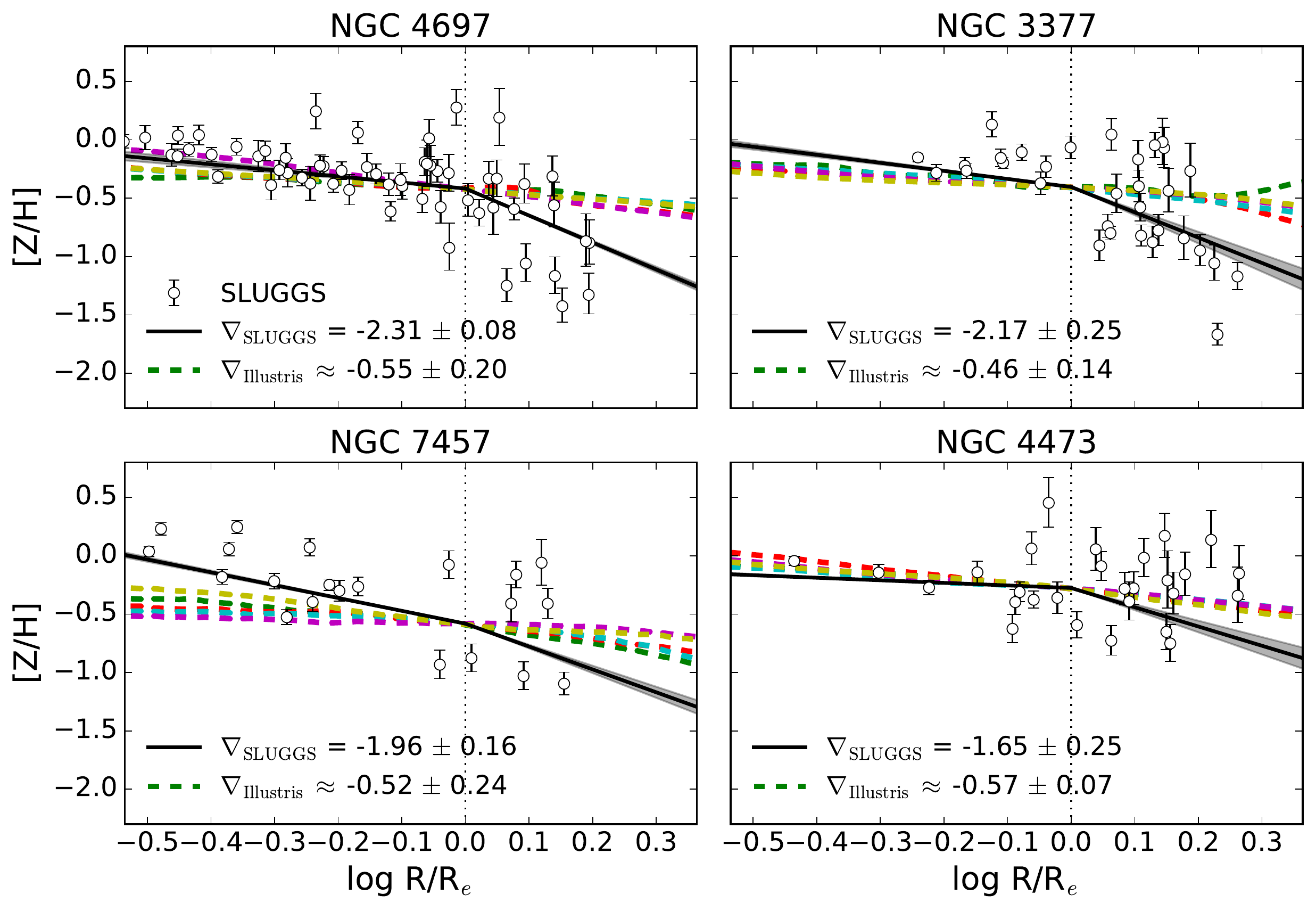}{0.95}
  \caption{Comparisons between the metallicity profiles of Illustris
    galaxies and the SLUGGS survey \citep{Pastorello2014} for the four
    SLUGGS galaxies with the steepest metallicity gradients. The
    measured metallicity data points are shown in white (Pastorello,
    private correspondence), while the profiles for a random selection
    of Illustris galaxies at similar masses are shown as dashed
    lines. The black lines denote profiles with the published outer
    gradients ($> 1 \Reff$) and accompanying uncertainties for each
    galaxy, normalized to the typical metallicity of the data points
    around $\mathrm{R}\approx \mathrm{R}_e$. While the Illustris
    profiles are too shallow to match all the observed data points,
    the disagreement with the computed gradients is not as substantial
    as would be suggested by Figure \ref{fig.obs}.}
  \label{fig.SLUGGS}
\end{figure*}

In \S\ref{s.compare} we showed our measurements of stellar population
gradients in Illustris galaxies, and found they agreed well with most
available observations from the literature. A notable exception
was in the outer galaxy ($1 - 2\;\Reff$) metallicity gradients, which
at low masses are much flatter than observations from the SAGES Legacy
Unifying Globulars and GalaxieS (SLUGGS)
survey \citep{Pastorello2014}. In Figure \ref{fig.SLUGGS}, we directly compare
the metallicity profiles of Illustris galaxies to the four SLUGGS
galaxies with the steepest published gradients.

The individual SLUGGS metallicity measurements are shown in white
(Pastorello, private correspondence), each along with example
metallicity profiles from a selection of Illustris galaxies with
comparable masses. Representative profiles with the published SLUGGS
metallicity gradients and corresponding uncertainties are overplotted
with a gray band.

While in the cases of NGC 4697 the Illustris profiles are too shallow
to match the observed data, they appear broadly consistent with the
data for NGC 3377, 4473, and 7457, despite the fact that the computed
gradients seem so different. Further exploration of these
discrepancies is beyond the scope of this work.

\bibliography{StellarHalos}

\end{document}